\documentclass[usenatbib, useAMS, a4paper]{mn2e}
\usepackage{times}
\usepackage{color}
\usepackage{xspace}
\usepackage{multirow}
\usepackage{graphicx}
\usepackage{amsmath}
\usepackage{natbib}
\usepackage{myaasmacros}
\usepackage[english]{babel}
\usepackage{bbold}
\usepackage[breaklinks, colorlinks, citecolor=blue, linkcolor=black, urlcolor=black]{hyperref}
\hypersetup{breaklinks=true, linktocpage=true}

\newcommand{\eg}{\mbox{e.\,g.}\xspace} 
\newcommand{\ie}{\mbox{i.\,e.}\xspace}

\newcommand{\wrt}{\mbox{w.\,r.\,t.}\xspace}

\newcommand{\Msol}{\mathrm{M}_{\odot}}

\def\lsim{\mathrel{\rlap{\lower3pt\hbox{$\sim$}}
    \raise1pt\hbox{$<$}}}                
\def\gsim{\mathrel{\rlap{\lower3pt\hbox{$\sim$}}
    \raise1pt\hbox{$>$}}}                
\newcommand{\tens}[1]{{\bf \sf #1}}
\renewcommand{\vec}[1]{\bmath{#1}}

\begin{document}
\title{Genetically modified halos: towards controlled experiments in $\Lambda$CDM galaxy formation}
\author[Roth et al]{Nina Roth, Andrew Pontzen, Hiranya V. Peiris \\
  Department of Physics and Astronomy, University College
  London, Gower Street, London, WC1E 6BT, UK \\
  Emails: n.roth@ucl.ac.uk, a.pontzen@ucl.ac.uk, h.peiris@ucl.ac.uk}

\date{Accepted ---. Received ---; in original form ---}
\maketitle

\newcommand{\ap}[1]{{\color{red} #1}}
\definecolor{darkgreen}{rgb}{0,0.7,0}
\definecolor{orange}{rgb}{1.0,0.4,0}
\definecolor{mypink1}{rgb}{0.858, 0.188, 0.478}
\newcommand{\nr}[1]{\textcolor{darkgreen}{NR: [#1]}}
\newcommand{\hvp}[1]{\textcolor{mypink1}{HVP: [#1]}}
\newcommand{\app}[1]{\textcolor{orange}{APP: [#1]}}

\newcommand{\C}{{\ensuremath\tens{C}}}
\renewcommand{\P}{{\tens{P}}}
\newcommand{\deltan}{{\vec{\delta}_n}}
\newcommand{\deltazero}{{\vec{\delta}_0}}
\begin{abstract}
  We propose a method to generate `genetically-modified' (GM) initial conditions for high-resolution simulations of galaxy formation in a cosmological context. Building on the Hoffman-Ribak algorithm, we start from a reference simulation with fully random initial conditions, then make controlled changes to specific properties of a single halo (such as its mass and merger history). The algorithm demonstrably makes  minimal changes to other properties of the halo and its environment, allowing us to isolate the impact of a given modification. As a significant improvement over previous work, we are able to calculate the abundance of the resulting objects relative to the reference simulation.
  
Our approach can be applied to a wide range of cosmic structures and epochs; here we study two problems as a proof-of-concept. First, we investigate the change in density profile and concentration as the collapse time of three individual halos are varied at fixed final mass, showing good agreement with previous statistical studies using large simulation suites. Second, we modify the $z=0$ mass of halos to show that our theoretical abundance calculations correctly recover the halo mass function. The results demonstrate that the technique is robust, opening the way to controlled experiments in galaxy formation using hydrodynamic zoom simulations.  

\end{abstract}

\section{Introduction}
\label{sec-Intro}
Understanding galaxy formation requires us to take account of the variety of halo assembly histories that build today's population. Many pressing questions --- such as the origin of varying morphologies \cite[e.g.][]{2013ApJ...771L..35V,2015ApJ...803...26P} and bulge sizes \citep{2015arXiv150403330K} --- will be answered by understanding the interplay between complex, non-linear physics and the various histories for mass accretion. The fundamental difficulty is that these histories are in turn determined by the random initial conditions seeded in the early universe.

This paper is the first in a series to directly tackle that problem using a novel approach. The most typical solution is to simulate large numbers of galaxies in a representative volume \cite[e.g.][]{2014MNRAS.445..175G,2015MNRAS.446..521S,2015MNRAS.448.3391C}. However, this is computationally expensive and limits the resolution that can be achieved for any single object. Conversely, zoom-in simulations achieve the maximum level of physical detail for a given computational time. They have been used to establish that qualitatively different processes come into play at sub-kpc resolutions, where  processes within the interstellar medium begin to be resolved \cite[e.g.][]{2007MNRAS.374.1479G,2011ApJ...742...76G,2011MNRAS.415.1051B,2013MNRAS.430.1901H,2014Natur.506..171P}.   But such approaches only sample over a small and potentially biased range of merger histories. A third tactic is to use isolated, idealised set-ups to test particular hypotheses \cite[e.g.][]{1999ApJ...523L.133N, 2006ApJ...645..986R, 2009ApJ...691.1168H}; but these by definition lack a full cosmological environment. Thus, it is difficult to quantitatively connect the results of isolated and zoom simulations to the observed galaxy population.

Our aim is to combine the best aspects of these three types of numerical study. We proceed by systematically changing aspects of individual galaxies' histories (such as mass and merger history) within a cosmological simulation, while keeping track of the statistical likelihood of the changes to understand the relative abundance of objects of different types.
 This can be achieved by using the Hoffman-Ribak algorithm (\citealt[][hereafter HR91]{HR91}; see also \citealt{1986ApJ...304...15B, Bertschinger87, 1996MNRAS.281...84V} for further theoretical background). A more common use for HR91 is to obtain simulations resembling the local universe by turning a given observational dataset (\eg the local distribution of galaxies) into a prescription for the initial conditions of a numerical simulation \citep{Bistolas1998, Mathis02,Kravtsov2002,Klypin2003, Hess13,2013MNRAS.432..894J,Sorce14}. There is a significant literature that uses this technique to  study the formation history of the Local Group \citep{Zavala09, Klimentowski2010, Libeskind2010, Iliev11, CLUESForeroRomero11,  Kitaura13, Doumler13, CLUESDayal13, CLUESNuza14, CLUESBrook14}.

Instead, we propose to use constrained initial conditions as an experimentation toolkit for the formation of a particular halo embedded in a cosmological volume. This approach has precedent: for example, \cite{Frenk99} used the HR91 method to create galaxy cluster initial conditions for a comparative study of numerical simulation codes. More recently, \cite{RomanoDiaz06, RomanoDiaz07} and \cite{Hoffman07} simulated a single dark matter object of $\sim 10^{12} M_{\odot}$ with different substructures to understand the impact of quiescent and violent accretion phases on the inner properties of the halo, and the origin of the universal halo density profile. By including baryonic physics, \cite{RomanoDiaz2011a, RomanoDiaz2011b, RomanoDiaz2014} studied galactic properties in extremely overdense regions which may host the early precursors of QSOs. In a similar vein, \cite{Dubois12} use the numerical implementation from \cite{Prunet08MPGRAFIC} to investigate the accretion of material in the cores of very massive halos to shed light on the formation of black holes at high redshifts.

In all of the above cases, a simulated object was created by constraining the properties of a region defined by an analytical profile (typically a Gaussian peak). Constrained properties included the height of the density peak at the origin and its first- and second-order derivatives (see \eg \citealt{Prunet08MPGRAFIC} and the Appendix of \citealt{RomanoDiaz2011a}). This creates objects that are well-defined in a theoretical sense (\eg one can predict their collapse time reasonably well), but that represent configurations which may or may not be common in fully random initial conditions (ICs).

What sets our work apart from these previous efforts is that we always start with a `reference' halo from a simulation based on fully random ICs. We are able to impose constraints on volumes of completely arbitrary shape, using the particles that make up a single dark matter halo embedded in a cosmological volume. Once the constraints are applied, we re-run the simulation and compare the results to the original reference run.

This has two immediate benefits. First, we can fine-tune selected properties of the halo while demonstrably ensuring that the constrained object is as similar as possible to the reference run --- a controlled `genetic modification' (GM) of the halo. Second, we can calculate the change in the likelihood of the field after the modification; in other words, we can assess the relative abundance of the genetically-modified systems compared to the original. This will allow us to test whether connections between merger history and morphology quantitatively account for observed population statistics.

The current work provides a first illustration of both these aspects of the technique. Specifically, we study the properties of several halos as their total mass and merger history are systematically changed. We investigate the concentration at $z=0$ for different mass accretion histories and find overall excellent agreement of our constrained halos with relations derived from statistical averages over large simulations. 
There are many studies that connect the concentration parameter to other halo properties like the mass, collapse time or mass accretion history, halo shape and angular momentum \citep[\eg][]{2001MNRAS.321..559B, 2002MNRAS.331...98V, 2002ApJ...568...52W, 2003MNRAS.339...12Z, 2005MNRAS.357...82R, 2007MNRAS.376..215B,  2007MNRAS.378...55M, 2007MNRAS.381.1450N, 2008MNRAS.390L..64D, 2008MNRAS.391.1940M, 2009ApJ...707..354Z, 2010MNRAS.407..581R,2012MNRAS.423.3018P, 2013MNRAS.432.1103L, 2014MNRAS.441..378L, 2014arXiv1411.4001K, 2014arXiv1409.5228C, 2015arXiv150104382C, 2015arXiv150200391C}. 
Often, these studies operate by considering a statistical sample from a large volume simulation to find correlations and provide fitting functions. Even though the statistical power in recent simulations is excellent, there is still considerable scatter around the median relations. Since the density profile of dark matter halos is an important ingredient in theoretical models, it is important to understand these correlations and the scatter. Given the large number of parameters that could influence the evolution of a halo, principal component analysis has been used to investigate correlations between them \citep{2011MNRAS.416.2388S, 2011MNRAS.415L..69J, 2012ApJ...757..102W}. 
Our approach of designing `experiments' in galaxy formation provides a complementary approach to computationally expensive statistical studies.

This paper is organised as follows: in Sec.~\ref{sec-tech} we give a brief outline of the HR91 technique and our specific implementation. Section \ref{sec-sims} contains details of the numerical simulations that are used to obtain the results in the rest of the paper. In Sec.~\ref{sec-masscon} we provide a brief illustration of some of the constraints we have applied to the reference initial density field, focusing on influencing a single halo traced by its particles. Next, we study the results of designing different merger histories for a set of halos in Sec.~\ref{sec-merger}, focusing on their collapse-concentration relation. In Sec.~\ref{sec-chisq} we discuss a method for assessing the relative abundance of the modified halos by defining a $\chi^2$ measure, and show that our results are consistent with the cosmological halo mass function. We summarise in Sec.~\ref{sec-conc}. Finally, Appendices~\ref{app-tech} and \ref{app-B} contain the mathematical details of our reformulation of the HR91 technique including a translation between our notation and theirs.

\section{Outline of the Method}
\label{sec-tech}
We now present a brief outline of the mathematical technique by which
initial conditions can be generated that satisfy certain constraints, while remaining consistent with a $\Lambda$CDM power spectrum. This technique is described in a slightly different formulation by HR91.
The full derivation can be found in Appendix~\ref{app-tech}.

By assumption, the density field in the early universe
is linearly perturbed around a background density $\rho_0$, so that
\begin{equation}
\rho(\vec{x}) = \rho_0 \left(1+\delta(\vec{x}) \right)\textrm{,}
\end{equation}
where $\delta(\vec{x})$ is a Gaussian random field with statistical properties specified by the $\Lambda$CDM transfer function and inflationary tilt.

Generating ICs involves sampling the Gaussian random field at a list of discrete points $\vec{x}_{\nu}$, where the integer value $\nu$ decides which point we are discussing. In particular, when running a uniform-resolution cosmological simulation with $N$ particles on a box side, $\nu$ runs from $1$ to $N^3$. The sampled field then consists of an $N^3$-length vector $\vec{\delta}$, where an element is given by $\delta_{\nu} \equiv \delta(\vec{x}_{\nu})$. 
The values of $\vec{\delta}$ are drawn from a multivariate Gaussian probability distribution with mean $\langle \vec{\delta} \rangle = \vec{\mu}_0$ and covariance matrix
$\tens{C}_0 = \langle (\vec{\delta}- \vec{\mu}_0 )^{\dagger} (\vec{\delta} - \vec{\mu}_0 ) \rangle $.
Cosmological initial conditions have zero mean, $\vec{\mu}_0 = 0$,
but the HR91 technique is not limited to this case. 

A constrained field is defined by requiring $\vec{\alpha}^{\dagger} \vec{\delta} = d$
for some constraint vector $\vec{\alpha}$, which also contains $N^3$ elements. In general, $d$ is real, though the formalism also extends to the case where it is complex-valued.
A simple example would be to fix the density contrast to 0 at position $\vec{x}_{1}$. 
This requires $\alpha_{\nu} = 1$ for $\nu=1$ (before normalisation, see below) and 0 otherwise, and $d=0$.
Throughout, we will use Greek indices in this way to denote values of either $\vec{\delta}$ or $\vec{\alpha}$ at a specific grid position $\vec{x}_{\nu}$.

To actually create a field satisfying any given constraint, one could sample repeatedly from the underlying population until obtaining a realization that satisfies (or is close to satisfying) the requirement. However, such an accept-reject algorithm would be computationally expensive to implement in practice; instead, the HR91 technique makes a mathematical rearrangement that requires only one set of random numbers to be generated. As detailed in Appendix~\ref{app-tech}, this rearrangement also shows that a constrained Gaussian random field remains Gaussian. This allows us to apply a large number of constraints independently, with the final result obeying
$\vec{\alpha}_i^{\dagger} \vec{\delta} = d_i$ for each $i$ where the Roman index $i$ denotes the $n$ different constraints. The properties of the constrained field are then determined by a new mean and covariance
\begin{align}
&\vec{\mu}_n = \vec{\mu}_0 + \sum_{i=1}^n \tens{C}_0 \vec{\alpha}_i \left( d_i-
\vec{\alpha}_i^{\dagger} \vec{\mu}_0\right) \\
&\tens{C}_n = \tens{C}_0 - \sum_{i=1}^n
\tens{C}_0\vec{\alpha}_i\vec{\alpha}_i^{\dagger} \tens{C}_0,
\label{eq-CN}
\end{align}
provided that the $\left\{\vec{\alpha}_i\right\}$ have been orthonormalized\footnote{Note that this orthonormalization can always be arranged for any set of non-conflicting original constraints.} in the sense that $\vec{\alpha}_i^{\dagger} \tens{C}_0 \vec{\alpha}_j = \delta_{ij}$.

\begin{table}
\begin{tabular}{|l|c|c|c|}
\hline
 Name & Halo & Constraint & Section  \\
 \hline
 Reference & N/A & none & \ref{sec-sims} \\
 H24-MH & Halo 24 & $d_{10}/d_{\mathrm{ref}} = \{\textbf{0.5}, 1.5 \}$ &\ref{sec-masscon}, \ref{sec-merger} \\
 H24-MH* & Halo 24 & $d(z=1)/d_{\mathrm{ref}} =\{0.9, 1.1 \} $  & \ref{sec-masscon}, \ref{sec-merger} \\
 H37-MH & Halo 37 & $d_{10}/d_{\mathrm{ref}} = \{\textbf{0.5}, 1.5 \}$ &\ref{sec-masscon}, \ref{sec-merger} \\
 H37-MH* & Halo 37 & $d(z=1)/d_{\mathrm{ref}} = \{0.9, 1.1 \}$ & \ref{sec-masscon}, \ref{sec-merger} \\
 H40-MH & Halo 40 & $d_{10}/d_{\mathrm{ref}} = \{0.5, 1.5, 2\} $ & \ref{sec-masscon}, \ref{sec-merger} \\
 H40-MH* & Halo 40 & $d(z=1)/d_{\mathrm{ref}} = \{\textbf{0.9}, 1.1, \textbf{1.2}\} $ & \ref{sec-masscon}, \ref{sec-merger} \\
 H24-mass & Halo 24 & $d/d_{\mathrm{ref}}=\{0.5, 0.8, 1.2, 1.5\}$ &\ref{sec-masscon}, \ref{sec-chisq} \\
 H37-mass & Halo 37 & $d/d_{\mathrm{ref}}=\{0.5, 0.8, 1.2, 1.5\}$ &\ref{sec-masscon}, \ref{sec-chisq} \\
 H40-mass & Halo 40 &  $d/d_{\mathrm{ref}}=\{0.5, 0.8, 1.2, 1.5\}$ &\ref{sec-masscon}, \ref{sec-chisq} \\
 \hline
\end{tabular}
\caption{Overview of the simulations used in this paper. For more details on the individual runs, 
see the text in the sections mentioned in the last column. MH and MH* stand for the two different ways of constraining the merger history of the halo.
The simulations marked in bold are not actually used because they are not in equilibrium at $z=0$ (see Sec. \ref{sec-merger} for a discussion).} 
\label{tab-sims}
\end{table}

A realization of the constrained field could therefore be obtained by calculating $\vec{\mu}_n$ and $\tens{C}_n$ and drawing random numbers accordingly. However, in practice, dealing directly with $\tens{C}_n$ from Eq.~\eqref{eq-CN} becomes prohibitively expensive for large $N$. The problem is that, whereas $\tens{C}_0$ is the $\Lambda$CDM power spectrum and therefore diagonal in Fourier space, $\tens{C}_n$ is generally not sparse in either pixel or Fourier space.
Instead, one can make the ansatz that a realization obeying $n$ constraints, $\vec{\delta}_n$, can be obtained starting from a realization of the unconstrained field, $\vec{\delta}_0$, via
\begin{equation}
\vec{\delta}_n = \tens{P}_n \left( \vec{\delta}_0 - \vec{\mu}_0
\right) + \vec{\mu}_n\textrm{,}
\label{eq-const}
\end{equation}
where $\tens{P}_n$ is a matrix that depends on $\tens{C}_0$ and $\left\{ \vec{\alpha}_i \right\}$.

By requiring that $\vec{\delta}_n$ obeys the correct statistics and, additionally, requiring that the changes made to the field are minimal, one can uniquely derive the HR91 solution for $\tens{P}_n$. The
details are given in Appendix \ref{app-tech}, with the result that
\begin{equation}
\vec{\delta}_n = \vec{\delta}_0 +\sum_{i=1}^n \C_0 \vec{\alpha}_i \left(d_i- d_{i0}\right),
\label{eq-dn-full}
\end{equation}
where we have defined $d_{i0} = \vec{\alpha}_i^{\dagger} \vec{\delta}_0$ to
represent the value of the constrained quantities in the unconstrained
realization, and again require that the $\left\{\vec{\alpha}_i\right\}$ are orthonormalized.
This reduces the actual calculation to a series of vector multiplications and summations in Fourier-space (since $\tens{C}_0$ is diagonal there). The memory requirements are managable since we need only store vectors of length $N^3$, instead of the $N^3 \times N^3$ matrix $\tens{C}_n$.

Consequently, a continuum of constrained realizations can be generated from a single realization of the original ensemble. We select a single dark matter halo from a reference run at $z=0$, and then return to the initial conditions and place constraints on the particles that make up this object. In this way our constraint regions are defined directly via the halo particles, without requiring any assumptions about the properties of density peaks in the initial conditions, or any type of direct smoothing of the density field (only indirectly through the halo finding at $z=0$).

\section{Simulation setup}
\label{sec-sims}
All our simulations were run with \textsc{P-Gadget-3} \citep{Springel2005, 2008MNRAS.391.1685S}. The initial conditions have been set up at redshift $z=99$ and evolved to $z=0$, saving 100 snapshots from $z=9$ equally spaced in scale factor. The cosmological model is $\Omega_{\mathrm{m}}=0.279$, $\Omega_{\mathrm{b}}=0.045$, $\Omega_{\Lambda}=0.721$, $\sigma_8=0.817$, $h=0.701$, $n_{\mathrm{s}}=0.96$, corresponding to a WMAP5 cosmology \citep{2009ApJS..180..306D}. While these cosmological parameters have been revised in more recent datasets, they allow an easier comparison of our results with the literature. 
All simulations have a (comoving) box size of $L=50\ h^{-1} \ \mathrm{Mpc} \sim 71.3\ \mathrm{Mpc}$, and $N_{\mathrm{part}}=256^3$ dark-matter particles, resulting in a particle mass of $8.24 \cdot 10^8 \Msol$. The Plummer equivalent force softening length (which limits the smallest accessible scales) is $\epsilon = 25.6$ kpc in comoving units.

We use the \textsc{subfind} code \citep{2001MNRAS.328..726S}, which finds halos with the friends-of-friends (FoF) method. \textsc{subfind} also identifies subhalos inside the top level FoF groups, but we always use the whole group for particle tracking. Each FoF group is assigned a unique number, sorted in descending order by mass. \textsc{subfind} also provides a list of halo particle IDs, which allows us to track the halos between snapshots and across different simulations (by selecting those objects which have the most particles in common at $z=0$). We choose the standard FoF linking length of 0.2 times the mean interparticle distance. 

Our analysis makes use of the Python module \textsc{pynbody} \citep{2013ascl.soft05002P}. We select halos with mass $M_{200} \sim 10^{13} \Msol$ at $z=0$, which are well-resolved but not the most massive (and therefore rare) objects in the box. Throughout this paper, $M_{200}$ refers to the halo mass contained in $r_{200}$, the radius within which the mass density is 200 times the critical density of the universe at that time. We construct halo merger trees by tracing halo particles from $z=0$ backwards in time through each simulation snapshot. This allows us to determine the mass accretion history and other internal properties of each object as a function of time. We take advantage of the fact that the density field is first set up by assigning one particle to each grid node (the displacements  are applied later). This means that any particle at $z=0$ can be traced back to a grid position $\vec{x}_{\nu}$ in the initial conditions, and no additional interpolation is necessary to calculate $\delta(\vec{x}_{\nu})$.

Table \ref{tab-sims} gives an overview of the different runs that are used in this paper; the last column provides the section where the constraints are described and the results are discussed. In order to show that the technique is robust, we have selected three halos of similar mass in the reference run, and constrained their properties in different ways. Therefore each simulation name contains a halo number and constraint type.

\begin{figure*}
\includegraphics[width=0.8\textwidth]{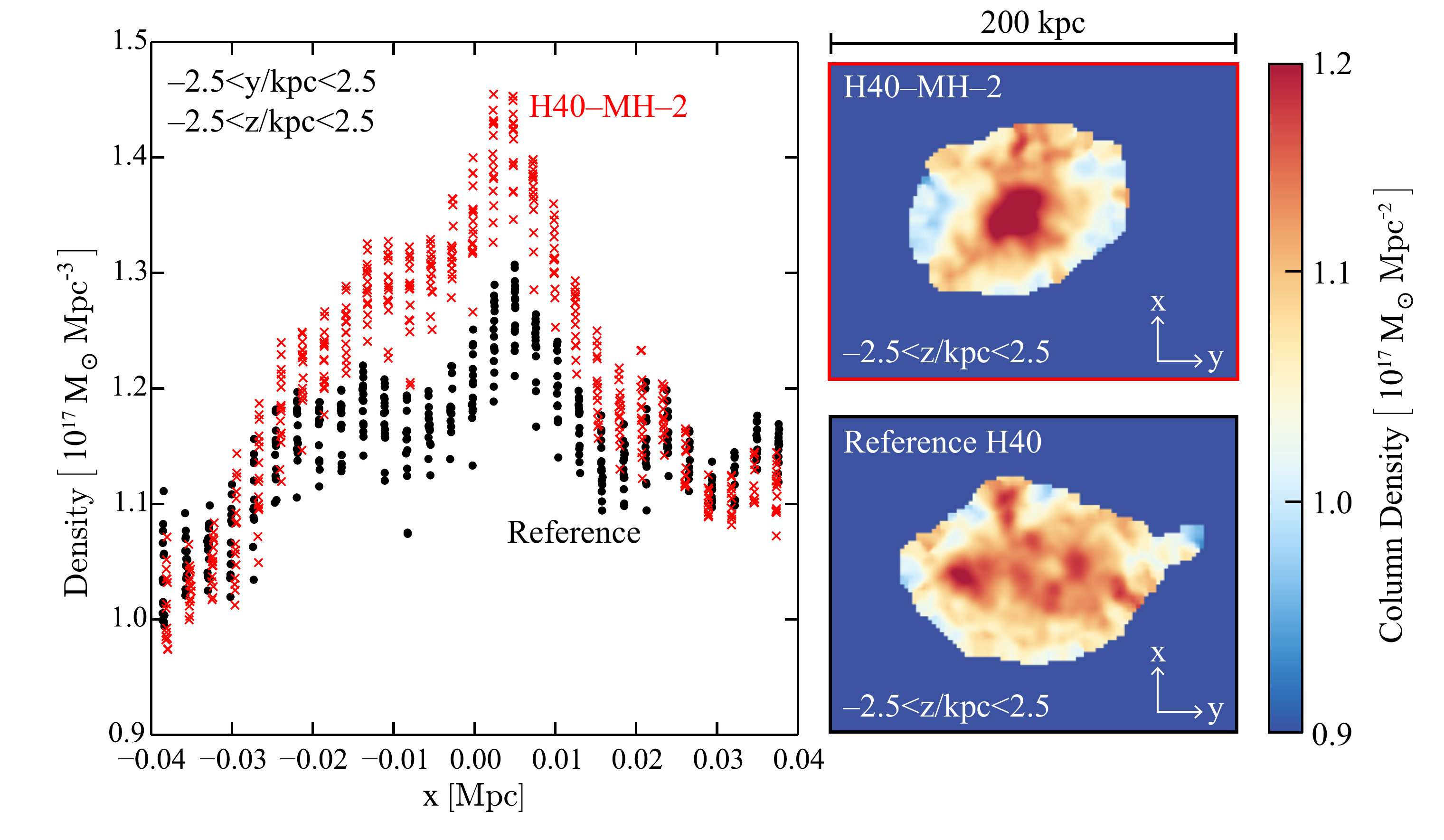}
\caption{
{\it Left panel}: The density of the reference ICs (black circles) and modified H40-MH-2 ICs (red crosses) for the early collapse constraint, where the density of the 10\% innermost particles is increased by a factor of 2. The slice is 5 kpc wide in the $y$- and $z$-coordinates, to give an impression of the 3D structure. Each symbol corresponds to a single particle/initial grid point. The constrained density field maintains the complicated (sub-)structure that was present in the reference run. {\it Right panels}: The same two ICs as a 2D projection in the $x-y$-plane. Only those particles that form each halo at $z=0$ are shown here; it is these particles that are used for generating the constraint in our algorithm. The higher central density is clearly visible in the constrained case. The results of these simulations will be discussed in detail in Sec.~\ref{sec-merger}.}.
\label{fig-1dmass}
\end{figure*}
\section{Illustration of constraints}
\label{sec-masscon}

We now present a simple illustration of the technique with which we generate a \emph{density constraint}. We will discuss the results obtained from running simulations with these constraints in the following sections.

Our approach constrains the actual Lagrangian region that collapses into a halo at $z=0$; by contrast, in previous work the constraints were typically chosen to follow some analytical form, in order to connect the constraints to theoretical models such as Press-Schechter theory (\eg \citealt{1996MNRAS.281...84V}, \citealt{RomanoDiaz06}). Since we know which particles are going to collapse in the reference run, we do not need to assume a specific form for the peak or a smoothing scale. The resulting constrained halo will be very similar to the reference object, unless the constraint radically changes the collapsing region, \eg by introducing a large overdensity in a region which only forms an intermediate mass halo in the reference run. 

Designing the constraints for a given modification to the final halo requires a physical understanding of the evolution. Ultimately a proposed constraint must be tested by trialling the changes and testing that they have the desired effect and that they are statistically consistent with the modified halo existing in the unconstrained universe. We will demonstrate both of these properties over the remainder of the paper.

Changing the mass can be achieved by changing the density contrast of the halo particles in the initial conditions. By creating a larger or smaller overdensity, we influence the final mass by increasing or decreasing the overall size of the region which has the average threshold density to collapse by a specified redshift \citep{1974ApJ...187..425P}. We term this a \emph{density constraint}: in the initial conditions, we calculate the average mass overdensity of all $N_{\mathrm{part}}$ particles in the reference halo
\begin{equation}
\frac{1}{N_{\mathrm{part}}}\sum_{{\nu}=1}^{N_{\mathrm{part}}} \delta(\vec{x}_{\nu}) \equiv d,
\label{eq-density-constraint}
\end{equation}
where we again use the fact that each particle corresponds to a grid position $\vec{x}_{\nu}$. Before orthonormalization, the value of the constraint vector $\vec{\alpha}$ is then $1/N_{\mathrm{part}}$ for each particle which belongs to the halo, and $0$ otherwise. The density can now be increased or decreased by enforcing the value of $d$. Results of simulations with different choices for $d$ that produce halos with higher or lower mass at $z=0$ will be used in Sec.~\ref{sec-chisq}.

More specifically, according to the Press-Schechter argument, the collapse time of a halo is related to its peak height $\nu=\delta / \sigma(R)$, where $\sigma(R)$ is the variance of the density field smoothed on a scale $R$. Therefore by fine-tuning the overdensity on different scales within the initial conditions we can modify the accretion history. In particular, by increasing (or decreasing) the density contrast in an \emph{inner region} of the halo, while requiring that the overall density contrast of the halo particles stays the same, we are able to generate a halo with very similar mass at $z=0$, but a faster (or slower) accretion history.

\begin{figure*}
\includegraphics[width=0.5\textwidth]{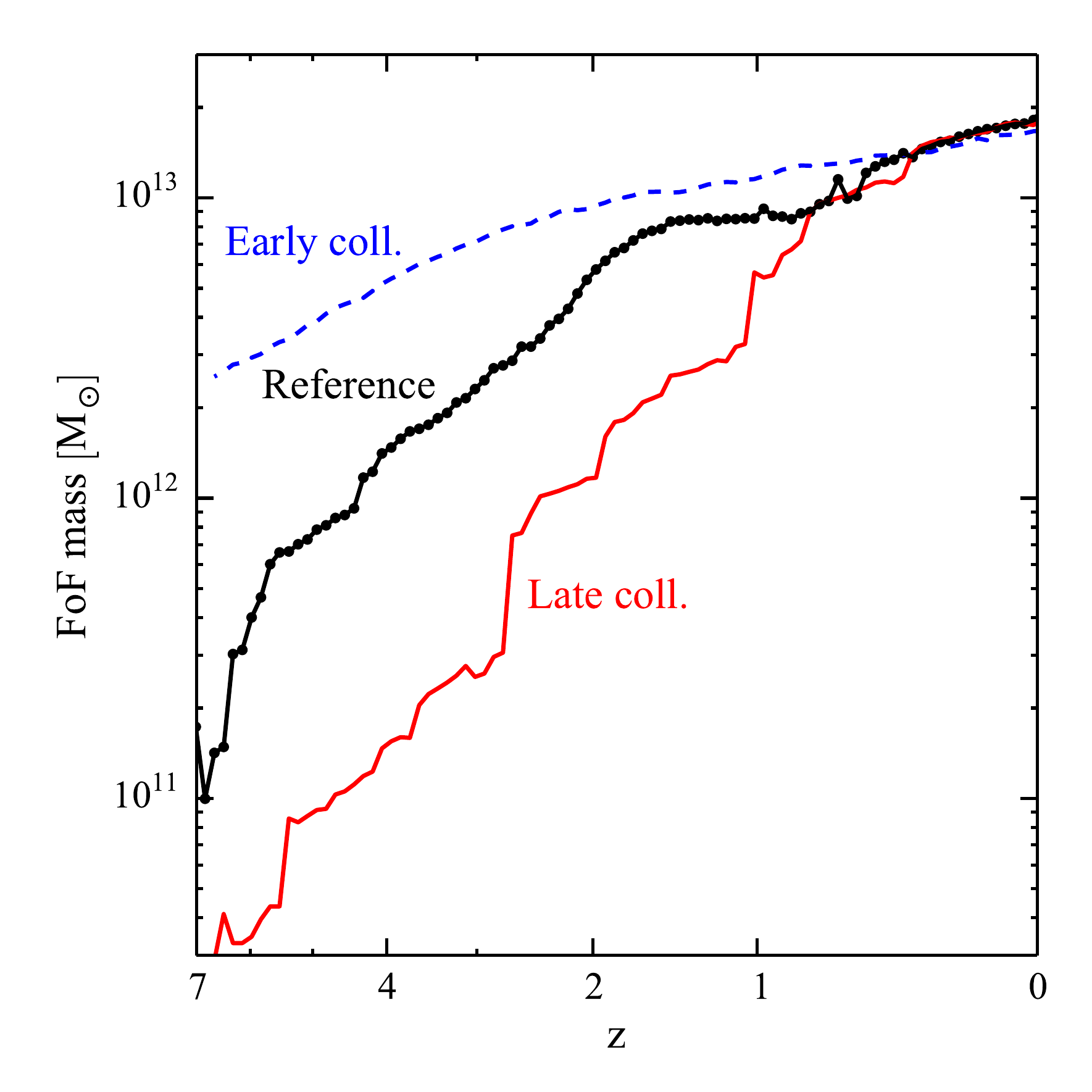}\includegraphics[width=0.5\textwidth]{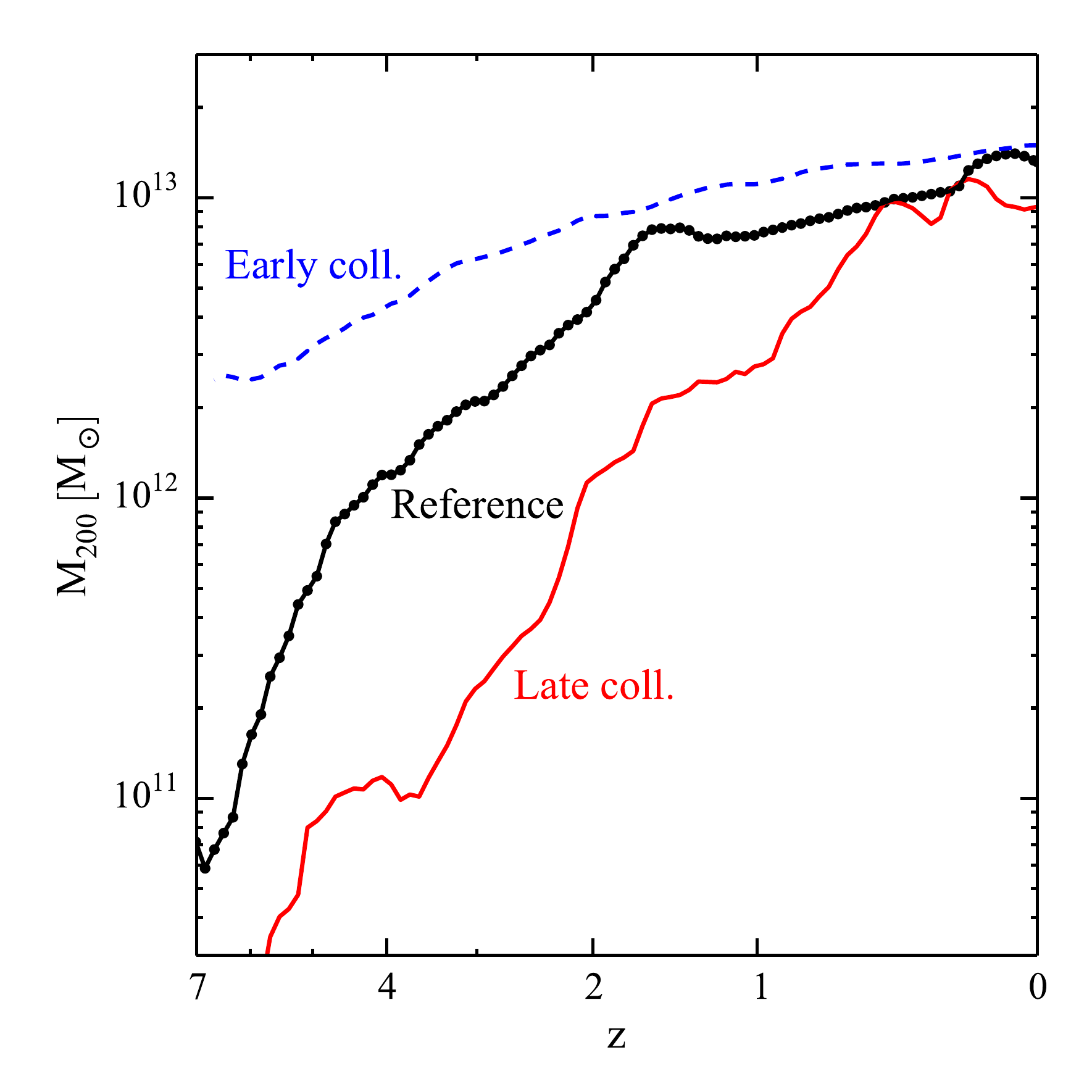}
\caption{{\it Left panel:} Mass accretion history for early (red solid) and late collapse (blue dashed) runs, expressed by the FoF mass (all particles assigned by the halo finder). The black solid line with points shows the same halo in the reference run; each point is one snapshot, illustrating the time resolution of our simulations. {\it Right panel:} same but for the virial mass $M_{200}$, which does not converge to a common value at late times because $M_{200}$ probes the inner regions of the halo (see Figure~\protect\ref{fig-densprof}), which are affected by the collapse time.}
\label{fig-masshist}
\end{figure*}

Figure \ref{fig-1dmass} illustrates an example of such a constraint acting on the initial conditions. In the left panel, we show a slice through the dark matter density field in the initial conditions, centred on the halo's centre of mass. Each black circle corresponds to a density value in the reference run, and the red crosses show the same position in the constrained run. The slice is 5 kpc wide in the $y$- and $z$-coordinates, to give an impression of the 3D structure.
Here, we show the constraint that will be used in the `early' run (discussed in detail in Sec.~\ref{sec-merger}), where we have increased the density of the innermost 10\% of halo particles to be a factor of two higher, while keeping the overall density the same. An alternative approach is to actually identify substructures at an intermediate redshift and apply the `inner' constraints to those specific particles.
We have tried this second approach for the present work, selecting the particles which have already collapsed around $z \sim 1$ by constructing a merger tree from the \textsc{subfind} output in the reference run.  Table \ref{tab-sims} contains an overview of all simulations used in this paper; we denote the first method of constraining the \emph{inner region} with $d_{10}$ and the second by $d(z=1)$ there. The two modification methods give near-identical results (Sec.~\ref{sec-merger}), making the outcomes reassuringly insensitive to the intuition guiding the modifications. We have found that it is also possible to add further constraints to modify the build-up in different subhalos and so fine-tune the accretion history to any required degree.

As explained in Sec.~\ref{sec-tech}, the modified field is constructed to follow the peaks and troughs of the underlying density field, thereby maintaining the same substructure as much as possible. In the right panel, we show the 2D projection ($x-y$-plane) of the density of halo particles in the initial conditions, again for both the reference run and constrained run. In both cases, the density is calculated for all particles that are part of the halo at $z=0$, which have been traced back to $z=99$. The effect of increasing the density in the innermost region can be clearly seen in the constrained run (red border). In addition the second constraint, which keeps the overall mass the same, leads to a compensation effect in the initial conditions, removing some particles in the outer regions which fall into the reference halo but not the constrained one. We will discuss the results of simulations with these modified initial conditions in the next section.

\section{Merger history and concentration-collapse relation}
\label{sec-merger}
So far we have looked at how applying various constraints modifies the initial linear overdensity field. We will now consider the changes that result when the new initial conditions are used in a numerical simulation, starting with our modified merger history.

The mass accretion histories for simulations with the H40-MH-2 `early' (circles) and H40-MH-0.5 `late' collapse constraint (crosses) are shown in Fig.~\ref{fig-masshist}. Here, we chose the innermost 10\% of particles and changed their overdensity by a factor of 2 (0.5) for the early (late) collapse cases. In the left panel we show the time evolution of the total mass (including all substructures) for the two constrained runs and the reference halo. It is clear that the accretion rates differ quite significantly at early times, but are compensated at late times, leaving the overall mass of the objects the same. In the right panel we show the time evolution of $M_{200}$. This quantity only measures the mass up to $r_{200}$ instead of the total mass of linked FoF particles (which can extend out to several $r_{200}$). As in the former case, the accretion rate follows the expected behaviour in the early and late collapse cases. However, at late times, $M_{200}$ differs between the constrained runs and the reference runs. This is due to a change in the halo density profile related to the collapse time, as we will now explore.

\begin{figure}
\includegraphics[width=0.5\textwidth]{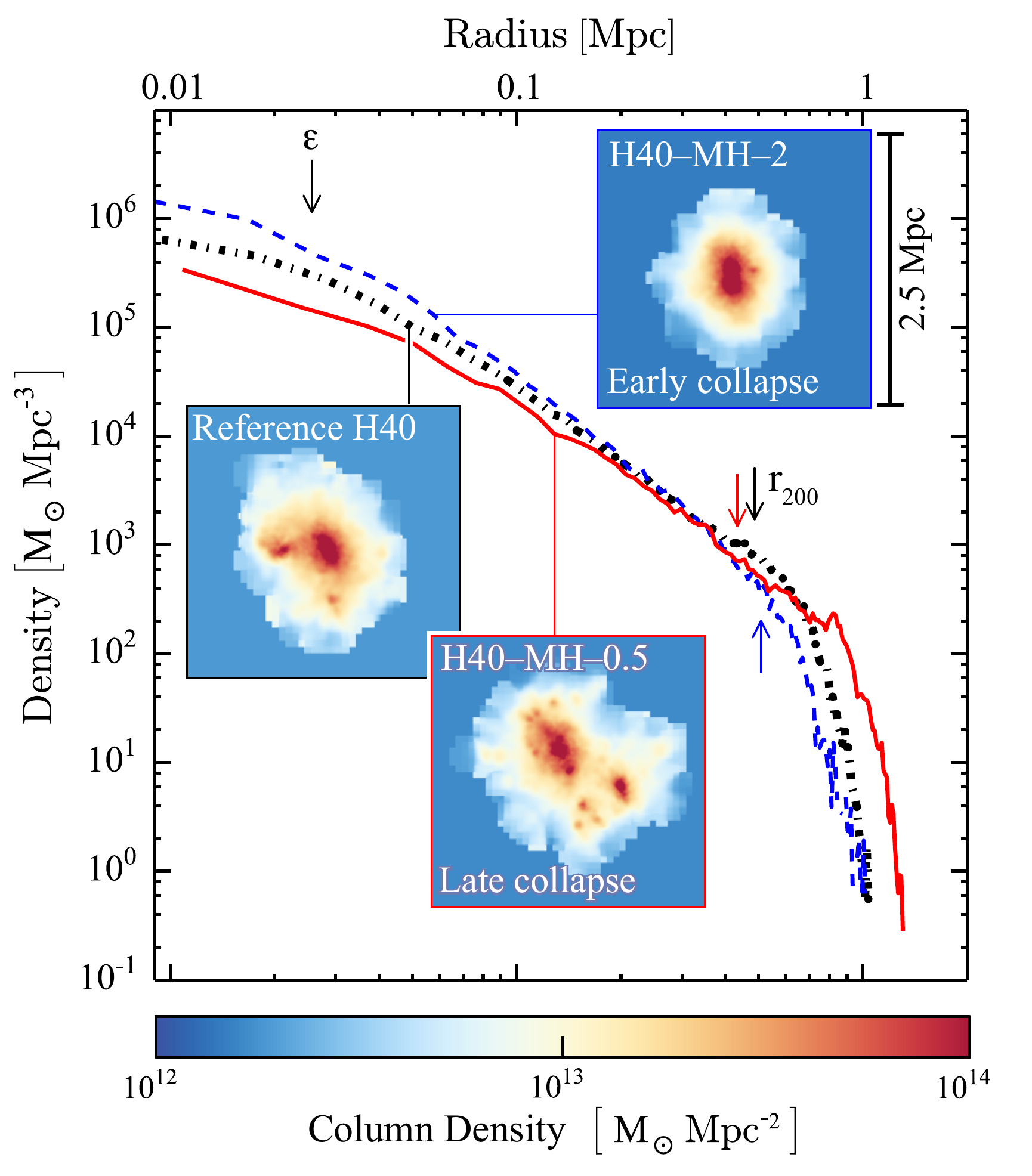}
\caption{Density profile of the reference halo (black dot-dashed) and the `early' (blue dashed) and `late' (red solid) constrained runs at $z=0$. The leftmost arrow indicates the softening length of the simulation, and the other arrows indicate the virial radius of each halo. {\it Inset panels:} density projection ($x-y$-plane) of the resulting halos at $z=0$. All panels show a region 2.5 Mpc across, include only the FoF group particles, and use the same colour scale for the column density.}
\label{fig-densprof}
\end{figure}

The halo radial density profiles for these three simulations at $z=0$ are illustrated in Fig.~\ref{fig-densprof}, with inset panels showing projected density maps. As the collapse is delayed, the slope in the inner regions becomes less steep. The location of the virial radius $r_{200}$ is indicated by an arrow in each case; as expected given our discussion above, this is displaced inwards by the relative shallowness of the late collapse H40-MH-0.5 case.

The difference between the density profiles can be encapsulated in the concentration parameter
\begin{equation}
c=\frac{r_{200}}{r_{\mathrm{s}}},
\label{eq-concparam}
\end{equation}
where $r_{200}$ is the virial radius of the halo (defined in Sec.~\ref{sec-sims}), and $r_{\mathrm{s}}$ is the scale radius in the NFW density profile \citep{1997ApJ...490..493N}
 \begin{equation}
\rho(r)= \frac{4 \rho_{\mathrm{s}}}{(r/r_{\mathrm{s}})(1+r/r_{\mathrm{s}})^2}.
\label{eq:defNFW}
\end{equation}
We fit an NFW profile to each of our halos at $z=0$, after determining its centre using a shrinking-sphere method and estimating the density in 100 radial bins of equal size. In order not to contaminate the fits with numerical artefacts introduced by the finite particle resolution, we exclude from the fit the innermost regions (2 times the softening length, $\epsilon$) which are affected by the force softening (\eg \citealt{power}), and regions with $r> 0.6\ r_{200}$, which may not be relaxed. We tested that the choice of the minimum and maximum radius has negligible impact on the estimate of $r_{\mathrm{s}}$. The measured values of the concentration are $4.2$, $6.1$ and $11.1$ for late, reference and early simulations respectively.

The GM method allows us to study the relationship between the collapse time of a halo (defined below) and its concentration as measured at $z=0$. Using a large statistical sample, \cite{2002ApJ...568...52W} found that $c \propto a_{\mathrm{coll}}^{-1}$, the scale factor at collapse time. We follow their procedure to obtain the collapse scale factor by fitting the mass accretion history of each halo with
\begin{equation}
M(z)=M_0\times \exp \left[-\alpha\ z\right],
\label{eq-massexp}
\end{equation}
with $M_0 \equiv M_{200}(z=0)$ and $\alpha= 2 a_{\mathrm{coll}}$. Extensions to this simple function have been proposed by \eg \cite{2004ApJ...607..125T}, \cite{2009MNRAS.398.1858M} and \cite{2014arXiv1409.5228C}, but for our purposes these are not necessary: the refined formulae are designed to accurately represent the median mass accretion histories for many halos, and the corrections are smaller than the scatter between individual halos.

\begin{figure*} \includegraphics[width=0.5\textwidth]{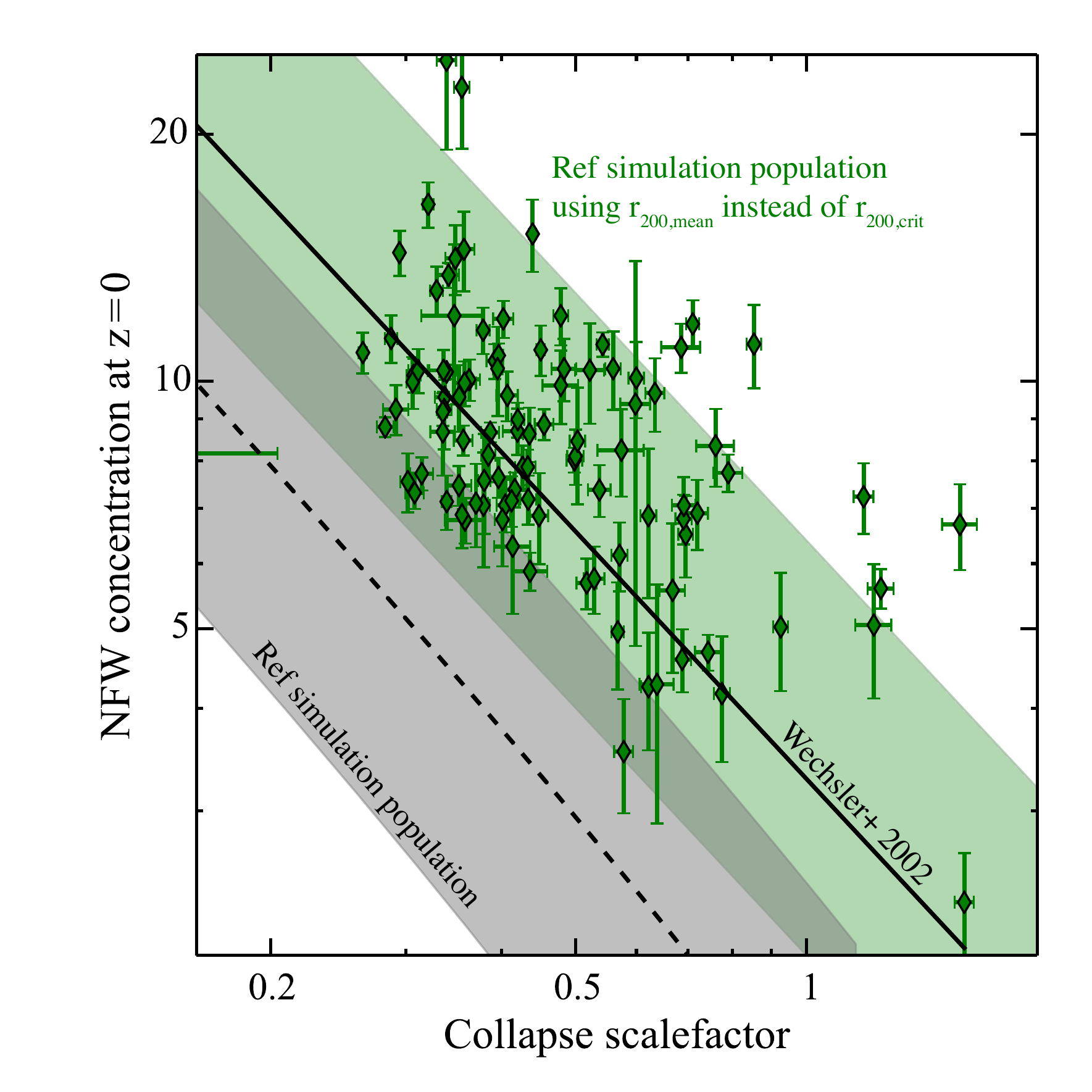}%
  \includegraphics[width=0.5\textwidth]{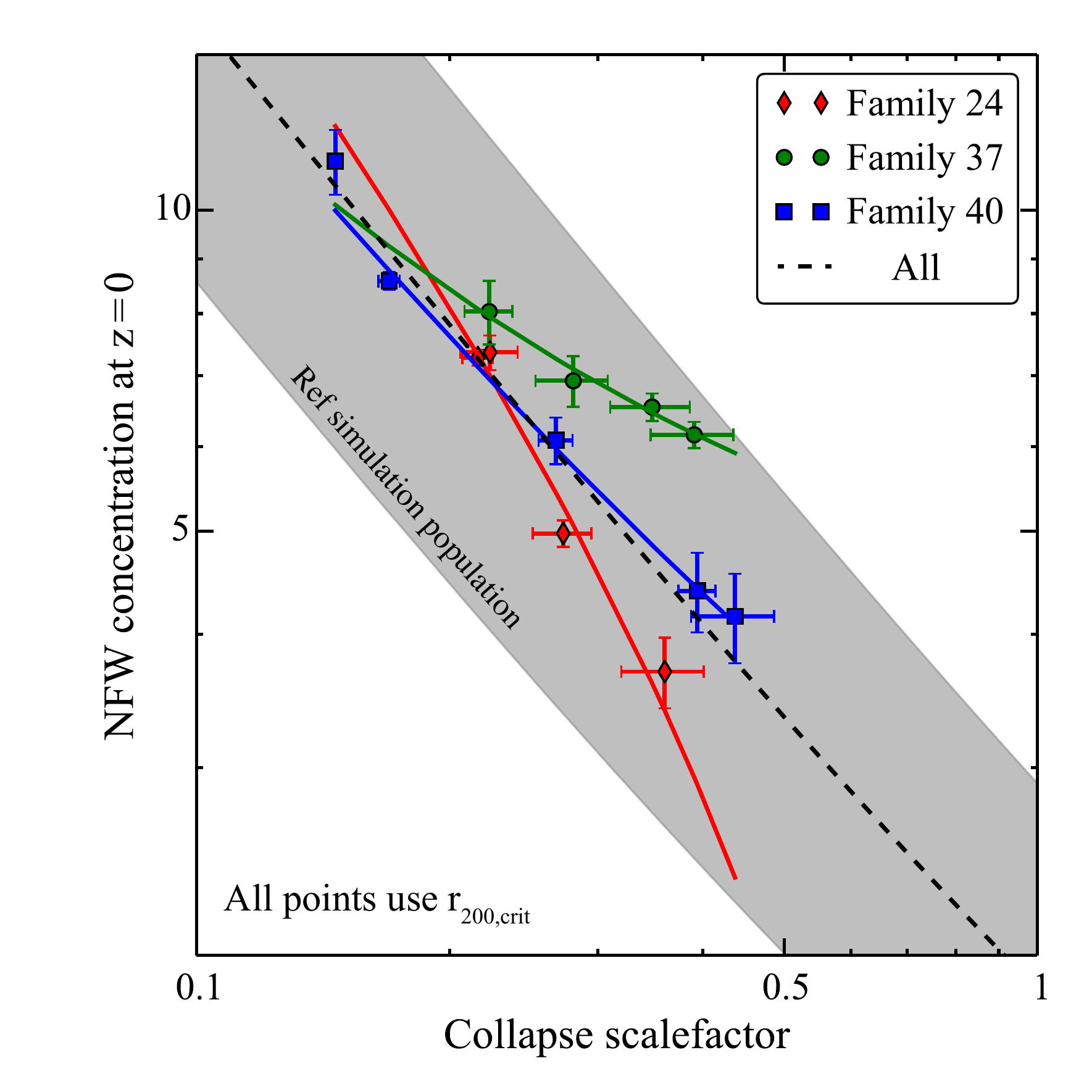}
  \caption{Halo concentration parameter as a function of the collapse time. {\it Left panel:} Our reference simulation gives a volume to probe the relationship using the traditional statistical technique. Taking 120 halos from our simulation, this results in a scatter of points in the region of the grey band. To compare with existing literature, we need to redefine $r_{200}$ (and hence $c$) relative to the mean (rather than critical) density, after which these halos are represented by the green points with error bars.  The black solid line and green band show the average relation and scatter as predicted by \protect\cite{2002ApJ...568...52W}, multiplied by a factor of 0.8 to account for their different choice of $\sigma_8$ \protect\citep{2008MNRAS.390L..64D}.  {\it Right panel:} Our three constrained halos (24, 37 and 40), showing fits to each constrained family individually (colours) and all of them together (black dashed). The grey band is the same as in the left panel. Together the panels establish that (left) halos in our reference volume recover the known relationship between concentration and collapse scale factor; and (right) relationships consistent with this relation are also recovered individually by each GM family. The scatter of slopes between different families is expected (see text).}
\label{fig-formcon}
\end{figure*}

There are some differences between the conventions of \cite{2002ApJ...568...52W} and the present work which we need to understand before proceeding. In the left panel of Fig.~\ref{fig-formcon}, the grey band and black dashed line show the average concentration and scatter from measuring the properties of $\sim 120$ halos of mass $M \sim 10^{13} M_{\odot}$ in our unmodified box. \cite{2002ApJ...568...52W} used a slightly different definition of halo mass from the one used in our study. Instead of defining $M_{200}$ with respect to the critical density of the universe, they define $M_{200}^{\mathrm{mean}}$ relative to the mean density.

To show how this affects the measured concentration, the green points show a sample of concentration parameters estimated using $M_{200}^{\mathrm{mean}}$, using the same halos that were used to generate the grey shaded region. There is an overall upward offset of these points relative to the grey band because $r_{200}^{\mathrm{mean}}$ is correspondingly larger. The black solid line and the green band show the median relation and scatter predicted by \cite{2002ApJ...568...52W} (taken from their Fig. 7), corrected by a factor of 0.8 following \cite{2008MNRAS.390L..64D} to account for their different $\sigma_8$ ($1$, instead of $0.817$ in our simulations).

In summary, once the differences in conventions and cosmological parameters are taken into account, we can reproduce the median results of \cite{2002ApJ...568...52W} in our unmodified boxes. The scatter from our simulation is also compatible with the much larger \cite{2002ApJ...568...52W} sample; outliers are likely due to the fact that we do not pre-select relaxed halos as they do. For consistency with the rest of the paper, we will use all quantities derived \wrt \emph{critical density} in the following analysis.

We are now ready to see how this relationship emerges when using GM halos instead of a statistical sample. For this study, we have selected three different halos in the reference run, which are all of similar mass ($M \sim 10^{13} M_{\odot}$).
The right panel of Fig.~\ref{fig-formcon} shows the result for 13 simulations (4 each for halos 24 \& 37, and 5 for halo 40) with different collapse times. We call each set a `halo family', including the reference run. The right panel of Fig.~\ref{fig-formcon} shows the results for the three families illustrated by red diamonds, green circles and blue squares for family 24, 37 and 40 respectively.

The slopes of the three families appear to be consistent with, but scattered around, the population average (grey band). We find that each halo family is well-described by a linear relation
\begin{equation}
c = \frac{\mathrm{const_1}}{a_{\mathrm{coll}}} + \mathrm{const_2},
\end{equation}
which contains the offset as an additional parameter compared to the fit used in \cite{2002ApJ...568...52W}. These fits are shown in the right panel of Fig.~\ref{fig-formcon} as coloured solid lines; we also fit all 13 points together (black dashed line). In addition, the grey band shows the scatter expected for halos in our selected mass bin, as in the left panel. The fit to all 13 simulations is very similar to the median relation from our unconstrained box which is consistent with the larger sample from \mbox{\cite{2002ApJ...568...52W}}.

The shift of each family member along its line is dictated by the direction and amplitude of the density constraint in the initial conditions. Higher values of the density in the inner region shift a point towards the top left, and lower values to the bottom right \wrt the reference run.  This gives us considerable insight into the kind of results that can be expected from GM compared to large population studies. The scatter of individual simulations within a GM family is very small --- in other words, the concentration is highly predictable from a single variable. This is because, as we have previously emphasized, the history of each halo within a single family is as similar as possible to all the others. The normal scatter in the concentration--collapse relation is then seen to be due to factors that are not being constrained within a single family (such as more detailed aspects of the merger history or other variables such as halo spin). The GM technique allows for a detailed exploration of results from specific, precise changes.

For halo 24 (red diamonds), two of the results are nearly identical: the point with the highest concentration value is actually two points nearly on top of each other. These points have been obtained using the two methods for setting mass accretion history constraints discussed previously, emphasising that they can lead to very similar results. Indeed, for each halo we have performed constrained runs using both methods, and a mixture of the resulting datapoints are shown in Fig.~\ref{eq-concparam}. For a list of all the simulations used in this paper, see Table \ref{tab-sims}.

The results of four additional runs (one each for halos 24 and 37, two for halo 40) are excluded from this Figure. In each case, the estimated density profile was not well-described by an NFW profile due to the halo undergoing a merger or the presence of large substructures. This is in agreement with \cite{2003MNRAS.339...12Z} who find that at least part of the scatter around the \cite{2002ApJ...568...52W} relation is due to poor fits to the NFW profiles and the mass accretion history.

\section{Likelihood of the modified field}
\label{sec-chisq}

As explained previously, the HR91 algorithm constructs a constrained realization
which is equivalent to (but much more efficient than) rejection sampling, \ie repeatedly drawing from
an ensemble of $\Lambda$CDM universes until one obtains a realization
that satisfies a given number of constraints. However, a naive choice of constraints can easily result in extreme configurations which are very unlikely to occur within the Hubble volume of the real universe. Depending on
context, this could even be intentional (\eg when investigating rare objects, \citealt{RomanoDiaz2011a, RomanoDiaz2011b, Dubois12, RomanoDiaz2014}); but nevertheless it is important to understand how likely it is for a given constrained configuration to arise, relative to the reference realization. We now derive a general expression for evaluating this likelihood and show how it is related to the abundance of halos when changing the mass. 

\begin{figure*}
\includegraphics[width=0.5\textwidth]{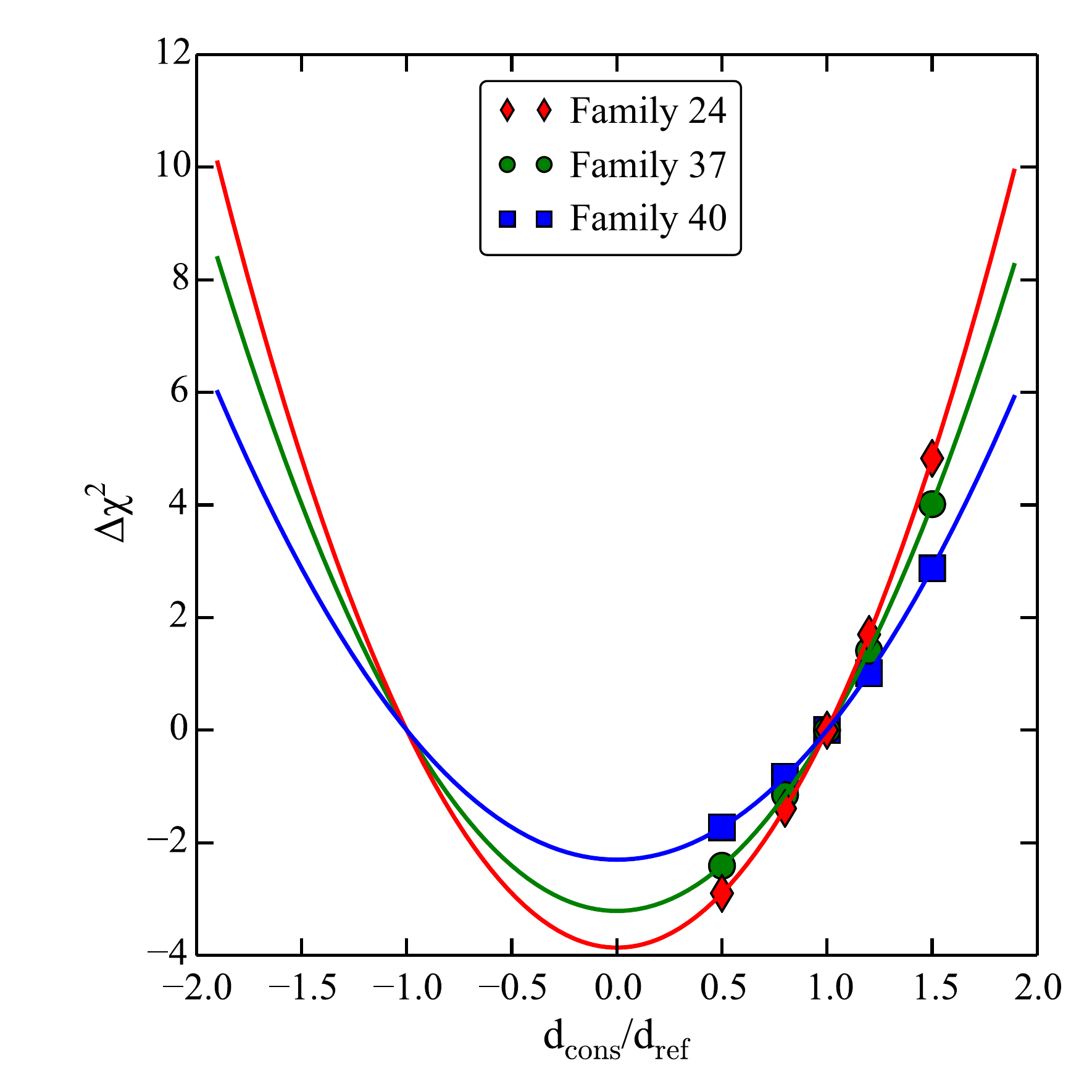}%
\includegraphics[width=0.5\textwidth]{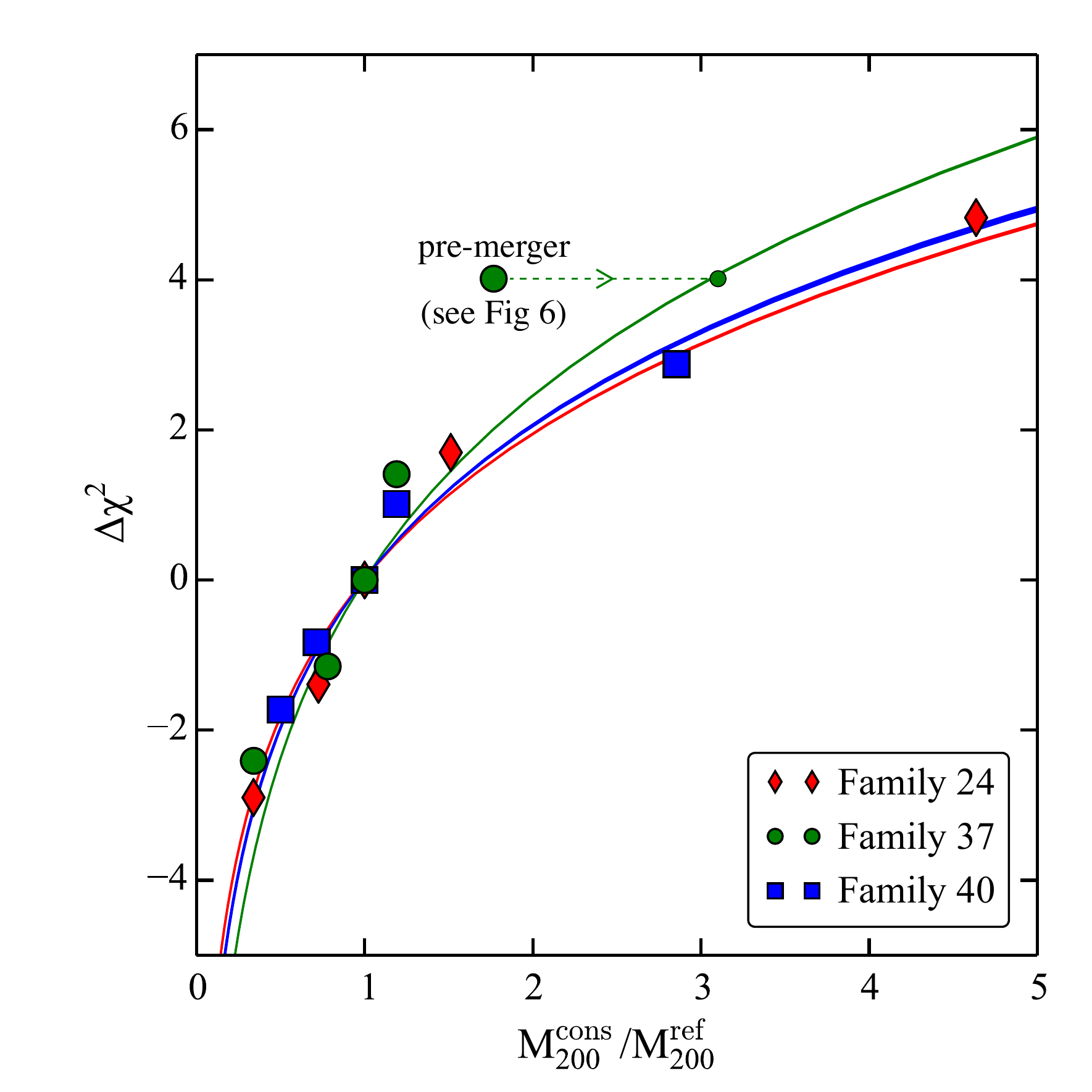}
\caption{{\it Left panel:} The relationship between $\Delta \chi^2$ and the initial overdensity for different halo families (H24-mass, H37-mass and H40-mass; see Table \ref{tab-sims}). Lines show the theoretical prediction from Eq.~\eqref{eq:dchi_theor}, whereas points give the actual change measured from the IC generator output, confirming that the algorithm is operating as expected. {\it Right panel:} $\Delta \chi^2$ values (points) can be interpreted as giving the relative abundance of the halos within each genetically-modified family, and therefore should agree with estimates from a halo mass function (lines). The agreement is indeed good except for the H37-mass-1.5 point which appears to have too small a mass at $z=0$ compared to expectations. This is because the halo mass function is based on an average mass build-up rate, whereas in this specific case the mass will only be acquired after a major merger in around $3\,\mathrm{Gyr}$ (see Fig.~\protect\ref{fig:halo37_1p5} and discussion in text).}
\label{fig-chisq}
\end{figure*}

We can compare the unconstrained and constrained fields
with respect to the \emph{unmodified} $\Lambda$CDM covariance matrix $\C_0$ by
evaluating the change in $\chi^2$, defined as
\begin{align}
\Delta \chi^2 &= \deltan^{\dagger}  \C_0^{-1} \deltan - \deltazero^{\dagger}
\C_0^{-1} \deltazero\textrm{,}
\label{eq-def-chisq}
\end{align}
where $\deltan$ is a field with $n$ constraints.
This constrained field has a relative abundance
in the universe of $e^{-\Delta \chi^2/2}$ compared to the original,
unconstrained field $\deltazero$. Since this is only a relative abundance, applying a constraint to a halo that is rare in the reference simulation will in general also generate a rare object in the constrained run. We therefore modify several halos in a similar way, in order to reduce the impact that picking a rare object may have on any of our results.

One can calculate $\Delta \chi^2$ directly from the density field, but we can also expand the above equation analytically by inserting Eq.~\eqref{eq-dn-full} and making use of $\vec{\alpha}_i^{\dagger} \C_0 \vec{\alpha}_j = \delta_{ij}$. This leads to a series of cancellations, with the final result
\begin{equation}
\Delta \chi^2 = \sum_{i=1}^n \left( |d_i|^2 - |d_{i0}|^2 \right) \textrm{,}
\label{eq:delta-chi2}
\end{equation}
where we again use $d_{i0} = \vec{\alpha}_i^{\dagger} \vec{\delta}_0$ to express the value of the constraints in the underlying realization.

Crucially, the details of the original realization $\vec{\delta}_0$ have
disappeared except in the initial values of the constrained quantities,
$d_{i0}$. In other words, the relative likelihood of the
constrained simulation compared to the unconstrained case is dependent
only on the choice of constraints. It is therefore specifically related to properties of the individual halo, not to details of its surroundings. This is another very desirable
property of the HR91 formalism and reflects the minimality
of the changes made to the field going from $\vec{\delta}_0$ to $\vec{\delta}_n$.

For a single constraint, Eq.~\eqref{eq:delta-chi2} has a
particularly transparent interpretation. Because of the normalization
condition, the variance of $d_{i0}$ for $i=n=1$ in unconstrained realizations is
\begin{equation}
\langle d_{0}^* d_{0} \rangle = \langle \vec{\delta}_0^{\dagger} \vec{\alpha}_1 \vec{\alpha}_1^{\dagger} \vec{\delta}_0 \rangle =
\vec{\alpha}_1^{\dagger} \tens{C}_0 \vec{\alpha}_1 = 1\textrm{.}
\label{eq:dchi_theor}
\end{equation}
Thus, for a single constraint, a change in $\Delta \chi^2$ of $1$ corresponds to a $1 \sigma$ variation in the property measured in the population-at-large.

The left panel of Fig.~\ref{fig-chisq} shows $\Delta \chi^2$ for a single constraint as a function of $d_{\mathrm{cons}}/d_{\mathrm{ref}}$, the ratio between a halo's average density contrast after the constraint and its value in the reference run (see Sec.~\ref{sec-masscon}). The values are calculated directly from the fields (points) and using Eq.~\eqref{eq:delta-chi2} (lines); the two methods agree to within numerical accuracy, which is a useful verification of the algorithm. As before, the results for the three families are illustrated by red diamonds, green circles and blue squares for haloes 24, 37 and 40 respectively. The minimum at $d_{\mathrm{cons}}=0$, as well as the symmetry, is expected for a zero-mean Gaussian random field.

\subsection{Connecting initial conditions and non-linear structure}
\label{ssec-massfun}
In this section, we work towards establishing a quantitative connection between the degree of change in the initial conditions and in the final, non-linear structure. Using a single constraint, we investigate how a change in density contrast in the initial conditions is related to the resulting halo mass at late times. Qualitatively, an increase in overdensity should lead to a more massive object at $z=0$ as explained in Sec.~\ref{sec-masscon}.

Quantitatively, the probability of finding a halo with mass $M$ at $z=0$ is given by the halo mass function, $n(M)\mathrm{d}M$, which depends on the cosmological power spectrum and growth function. Given two halos of mass $M_0$ and $M_1$, their relative abundance is given by the ratio of the halo mass function at those masses, $n(M_1)/n(M_0)$. Assuming that the statistical properties of constrained and unconstrained simulations can be related by the change in mass of the target halo alone, this ratio also gives the relative probability of the structure in the two simulations.

Additionally, we can calculate the relative probability of the two initial conditions using Eq.~\eqref{eq-def-chisq}; specifically
\begin{equation}
\frac{p(d_1)}{p(d_0)} = \exp\left[ - \Delta\chi^2/2 \right] ,
\label{eq-pdelta}
\end{equation}
where $p(d)$ is short-hand for the probability of a constrained field using one `density constraint' with value $d$. Since we only consider probability ratios and $\Delta \chi^2$, any terms which only change the normalization of $p(d)$ have dropped out.

Now we have two methods of calculating the relative probabilities, and we can check whether they agree. We rewrite $p(d)$ in terms of halo mass using the conservation of probability
\begin{equation}
p(M) = p(d)\, d'(M) ,
\label{eq-pmass}
\end{equation}
where $M$ is the halo mass at $z=0$, and $d'(M)$ is the derivative of the constraint $d$ with respect to the mass, evaluated at $M$. By combining Eqs.~\eqref{eq-pdelta}~and~\eqref{eq-pmass}, we can find a relationship between the $p(M_1)/p(M_0)$ and $\Delta \chi^2$; namely
\begin{equation}
\frac{p(M_1)}{p(M_0)} = \exp \left[ -\Delta \chi^2/2 \right] d'(M_1) \left[d'({M_0})\right] ^{-1}.
\label{eq-massfunratio}
\end{equation}
If our assumptions are correct, this expression for $p(M_1)/p(M_0)$ should be equal to the mass function ratio $n(M_1)/n(M_0)$. We have used HMFcalc \citep{2013A&C.....3...23M} to generate a halo mass function at $z=0$ for our cosmological model, and confirmed that it provides a good fit to our simulations.\footnote{This is preferable to obtaining a (noisy) estimate of the mass function directly from our limited volume; the fitting functions included in the HMFcalc tool were validated using detailed studies of large simulation suites \cite[\eg][]{2008ApJ...688..709T}.}

Evaluating Eq.~\eqref{eq-massfunratio} also requires an estimate of the Jacobian factors on the right hand side. This can be obtained either from a physical model underlying the mass function or by using an empirically calibrated $M(d)$ relationship from the simulations. We chose the latter approach by fitting a power-law relation between $d$ and $M$, which allows us to obtain values for $d'(M)$ at different halo masses separately for each halo family (24, 37 \& 40; introduced in the previous section). This leaves us with a `semi-analytical' prediction: theoretical halo mass function plus fit to the Jacobian.
\begin{figure}
\includegraphics[width=0.5\textwidth]{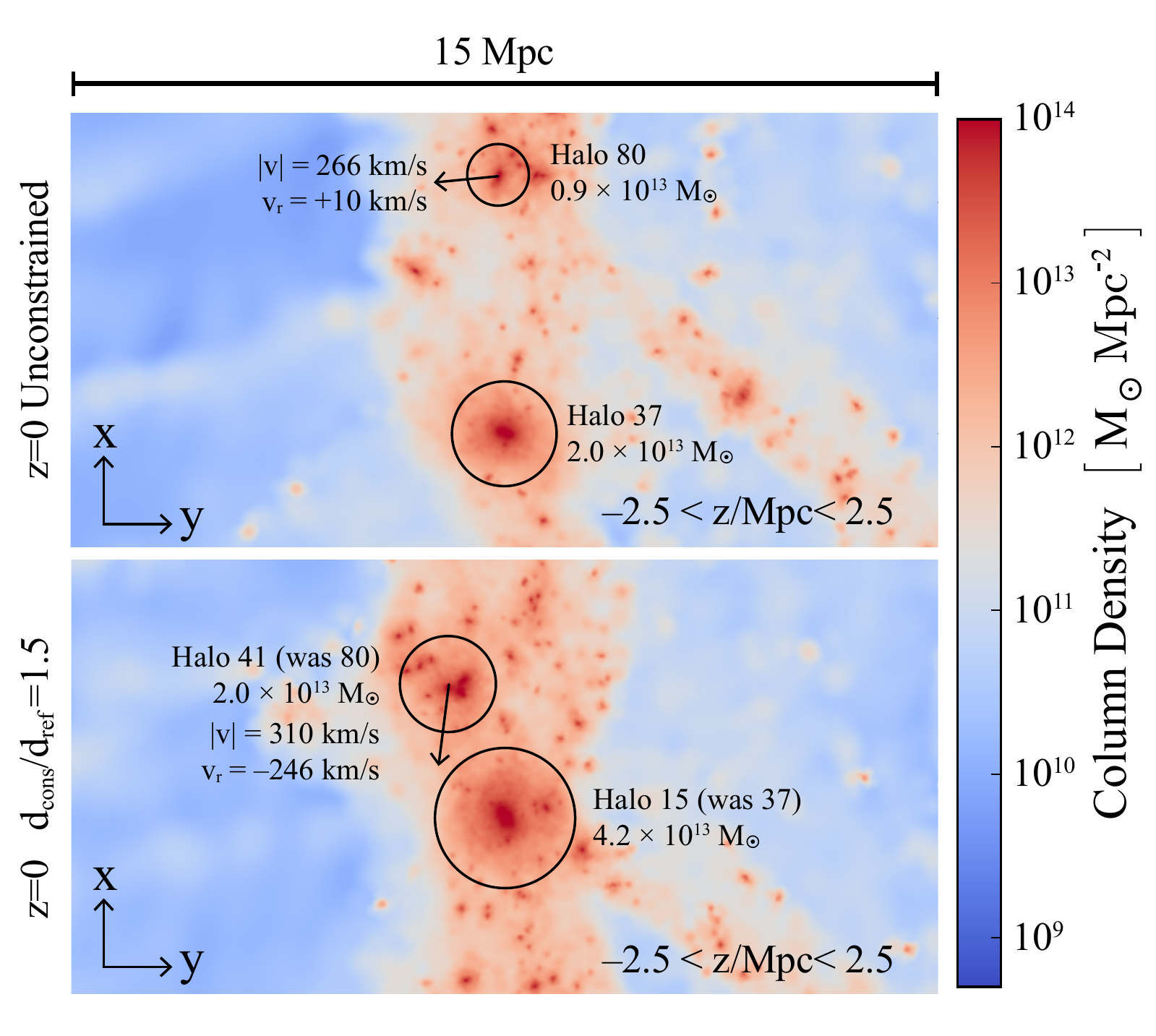}
\caption{Slices from the original simulation (upper panel) and H37-mass-1.5 simulation (lower panel) illustrate how the target halo is, at $z=0$, seen at a time where it is about to undergo a major merger in the latter case. For that reason its mass undershoots the expectation from the halo mass function (Fig.~\ref{fig-chisq}) which averages over all the possible discrete realizations of the accretion history. Black circles show the size of the virial radius.}
\label{fig:halo37_1p5}
\end{figure}

The right panel of Fig.~\ref{fig-chisq} shows the results of the calculation (lines), as well as points evaluated directly from the simulations. Overall, most points show a good agreement: there is consistency between the population statistics and the abundance calculated from the GM $\Delta \chi^2$ values. The broad agreement justifies our set of assumptions for calculating abundances in this specific case of a single density constraint. However the individual halos do scatter around the relation and there is one point that clearly does not fit the expectations. This arises from our H37-mass-1.5 run where we increase the initial overdensity of the proto-halo 37 region by a factor of 1.5.

The mismatch can be understood by considering the discrete nature of merger histories. Specifically, Fig.~\ref{fig:halo37_1p5} shows the projected density at the last output ($z=0$) in a region around halo 37 in the original run (upper panel) and the $\times 1.5$ run (lower panel). In the latter case, a major merger (mass ratio $\sim 2$) will occur in around $3\,\mathrm{Gyr}$. After this merger, the anomalous point will shift significantly rightwards in Fig.~5 to the correct mass ratio ($\simeq 3.1$) according to the $\Delta \chi^2(M)$ derived from the halo mass function.\footnote{Note that the nearest massive halo in the original run corresponds to the same particles, but is considerably further away from halo 37 and is not on a trajectory that will lead to a merger within a Hubble time.} We can frame this in another, more general way: the halo mass function is a statistical construction that corresponds to averaging over all possible histories, but the individual points in Fig.~\ref{fig-chisq} represent modifications to specific halos which have a discretized accretion history. Therefore, they scatter away from the line, especially when seen at special times (such as shortly before a major merger).

The main conclusion from Fig.~\ref{fig-chisq} is therefore that the changes in the $\chi^2$ give us a good quantitative handle on the relative abundance of halos of different types, at least in this case where we have only changed the mass. However, the complex non-linear connection between initial and final states means that $\Delta \chi^2$ will always need to be interpreted with care.

\section{Discussion \& Conclusions}
\label{sec-conc}

In this paper we have demonstrated an extension of the HR91 technique to modify the initial conditions of a numerical simulation. For this modification, we selected regions of arbitrary shape, defined solely by the particles that form a halo in our reference simulation. This is a different approach than that used in previous works, which relied on imposing constraints of a given analytic profile. By applying our constraints only to the halo particles, we showed that we can `genetically modify' a single object, changing its properties in a smooth and continuous way.

Using constraints on the density averaged over all halo particles controls the total mass, whereas adding additional constraints allows us to change the halo's collapse time. This serves as a demonstration of the technique and is the basis of the creation of further constraint types to study the impact of other halo characteristics on a halo's evolution.

Using the collapse time constraint, we investigated the density profiles of the resulting halos at $z=0$ and found that the distribution of their concentration parameter is consistent with the results of statistical analyses such as \cite{2002ApJ...568...52W}. However, we also find that different halos occupy different regions in the parameter-space, and have different trajectories when their collapse time is changed. We plan to study this behaviour in future work, in order to determine which other halo parameters have changed. This should be complementary to the principal component analysis carried out by \cite{2011MNRAS.416.2388S, 2011MNRAS.415L..69J, 2012ApJ...757..102W}, which revealed somewhat inconclusive correlations between additional internal halo parameters. With our constrained simulations, we will not only be able to find correlations but to explicitly test their significance. Since we can directly compare the constrained halo to its reference in the unconstrained run, we can establish exactly which changes in the halo parameters have a physical impact.

We have provided a way of quantifying the likelihood of modified initial conditions via a $\chi^2$ difference between the constrained and unconstrained fields. In general, this statistic can be used to assess how compatible the modified object is with the underlying cosmology. Similar expressions were obtained by \cite{1996MNRAS.281...84V}; however our orthonormalisation procedure allows for the derivation of the considerably simpler Eq.~\eqref{eq:delta-chi2}. The statistic can be used to quantify the rarity of genetically modified objects relative to the unconstrained realization. As an example we showed that modifying the mass produces abundance constraints that are quantitatively consistent with the traditionally-measured halo mass function at $z=0$. Individual halos have discrete accretion histories and scatter around the mean relation predicted by the mass function; the strongest outlier in our study is about to undergo a merger at $z=0$, which significantly lowers its current mass. In principle, the $\Delta \chi^2$ measure could also be used to specifically create objects that are `rare' in a $\Lambda$CDM universe \cite[similar to \eg][]{RomanoDiaz2011a, RomanoDiaz2011b, Dubois12, RomanoDiaz2014}; we leave such a study to future work.

In this paper we have used a uniform resolution over a box size of 50 $h^{-1}$ Mpc; having a sufficiently large box is important to ensure that halos are embedded in the correct large-scale environment. Our code also produces ICs for `zoom' simulations with varying resolution \cite[see also][]{Prunet08MPGRAFIC,RomanoDiaz2014}; the only major difference when generating these is the extra computational complexity introduced in the transformation between real space and Fourier space on an irregularly-spaced grid, which has been tackled elsewhere in the literature \citep[e.g.][]{2001ApJS..137....1B,2011MNRAS.415.2101H}. The real power of the approach to generate insight into a population from a handful of runs will become more apparent as we begin to use these zoom ICs in tandem with high-resolution baryonic physics.

While our focus here has been on basic properties such as the formation time and mass of a system, many other interesting aspects of evolution can be changed by constraining different properties. One example with which we are experimenting is the specific angular momentum, which can be controlled because tidal torque theory describes the connection between ICs and final spin \citep[e.g.][]{1984ApJ...286...38W,CatelanTheuns96,Porciani02a,Porciani02b}; the internal properties of the galaxy forming inside the dark matter halo will naturally depend on the spin parameter of the halo. We are able to generate initial conditions that modify the spin parameter of the halo, but leave the mass and merger history untouched. 
Studies of halo spin constraints for dark matter and hydrodynamic simulations will be presented in future work.

\section*{Acknowledgements}
We thank Volker Springel for allowing access to \textsc{P-Gadget-3} and \textsc{subfind}. NR thanks Emilio Romano-D{\'{\i}}az and Cristiano Porciani for useful discussions. AP acknowledges helpful conversations with Fabio Governato. NR and HVP are supported by STFC and the European Research Council under the European Community's Seventh Framework Programme (FP7/2007-2013) / ERC grant agreement no 306478-CosmicDawn. AP is supported by a Royal Society University Research Fellowship.
This work used the DiRAC Complexity system, operated by the University of Leicester IT Services, which forms part of the STFC DiRAC HPC Facility (\href{www.dirac.ac.uk}{www.dirac.ac.uk}). This equipment is funded by BIS National E-Infrastructure capital grant ST/K000373/1 and  STFC DiRAC Operations grant ST/K0003259/1. DiRAC is part of the National E-Infrastructure.

\bibliographystyle{mnras}
\bibliography{paper1refs}

\begin{thebibliography}{}
\makeatletter
\relax
\def\mn@urlcharsother{\let\do\@makeother \do\$\do\&\do\#\do\^\do\_\do\%\do\~}
\def\mn@doi{\begingroup\mn@urlcharsother \@ifnextchar [ {\mn@doi@}
  {\mn@doi@[]}}
\def\mn@doi@[#1]#2{\def\@tempa{#1}\ifx\@tempa\@empty \href
  {http://dx.doi.org/#2} {doi:#2}\else \href {http://dx.doi.org/#2} {#1}\fi
  \endgroup}
\def\mn@eprint#1#2{\mn@eprint@#1:#2::\@nil}
\def\mn@eprint@arXiv#1{\href {http://arxiv.org/abs/#1} {{\tt arXiv:#1}}}
\def\mn@eprint@dblp#1{\href {http://dblp.uni-trier.de/rec/bibtex/#1.xml}
  {dblp:#1}}
\def\mn@eprint@#1:#2:#3:#4\@nil{\def\@tempa {#1}\def\@tempb {#2}\def\@tempc
  {#3}\ifx \@tempc \@empty \let \@tempc \@tempb \let \@tempb \@tempa \fi \ifx
  \@tempb \@empty \def\@tempb {arXiv}\fi \@ifundefined
  {mn@eprint@\@tempb}{\@tempb:\@tempc}{\expandafter \expandafter \csname
  mn@eprint@\@tempb\endcsname \expandafter{\@tempc}}}

\bibitem[\protect\citeauthoryear{{Bardeen}, {Bond}, {Kaiser}  \&
  {Szalay}}{{Bardeen} et~al.}{1986}]{1986ApJ...304...15B}
{Bardeen} J.~M.,  {Bond} J.~R.,  {Kaiser} N.,   {Szalay} A.~S.,  1986, \mn@doi
  [\apj] {10.1086/164143}, \href
  {http://adsabs.harvard.edu/abs/1986ApJ...304...15B} {304, 15}

\bibitem[\protect\citeauthoryear{{Bertschinger}}{{Bertschinger}}{1987}]{Bertschinger87}
{Bertschinger} E.,  1987, \mn@doi [\apjl] {10.1086/185066}, \href
  {http://ukads.nottingham.ac.uk/abs/1987ApJ...323L.103B} {323, L103}

\bibitem[\protect\citeauthoryear{{Bertschinger}}{{Bertschinger}}{2001}]{2001ApJS..137....1B}
{Bertschinger} E.,  2001, \mn@doi [\apjs] {10.1086/322526}, \href
  {http://adsabs.harvard.edu/abs/2001ApJS..137....1B} {137, 1}

\bibitem[\protect\citeauthoryear{{Bett}, {Eke}, {Frenk}, {Jenkins}, {Helly}  \&
  {Navarro}}{{Bett} et~al.}{2007}]{2007MNRAS.376..215B}
{Bett} P.,  {Eke} V.,  {Frenk} C.~S.,  {Jenkins} A.,  {Helly} J.,   {Navarro}
  J.,  2007, \mn@doi [\mnras] {10.1111/j.1365-2966.2007.11432.x}, \href
  {http://adsabs.harvard.edu/abs/2007MNRAS.376..215B} {376, 215}

\bibitem[\protect\citeauthoryear{{Bistolas} \& {Hoffman}}{{Bistolas} \&
  {Hoffman}}{1998}]{Bistolas1998}
{Bistolas} V.,  {Hoffman} Y.,  1998, \mn@doi [\apj] {10.1086/305080}, \href
  {http://ukads.nottingham.ac.uk/abs/1998ApJ...492..439B} {492, 439}

\bibitem[\protect\citeauthoryear{{Brook} et~al.,}{{Brook}
  et~al.}{2011}]{2011MNRAS.415.1051B}
{Brook} C.~B.,  et~al., 2011, \mn@doi [\mnras]
  {10.1111/j.1365-2966.2011.18545.x}, \href
  {http://adsabs.harvard.edu/abs/2011MNRAS.415.1051B} {415, 1051}

\bibitem[\protect\citeauthoryear{{Brook}, {Di Cintio}, {Knebe},
  {Gottl{\"o}ber}, {Hoffman}, {Yepes}  \& {Garrison-Kimmel}}{{Brook}
  et~al.}{2014}]{CLUESBrook14}
{Brook} C.~B.,  {Di Cintio} A.,  {Knebe} A.,  {Gottl{\"o}ber} S.,  {Hoffman}
  Y.,  {Yepes} G.,   {Garrison-Kimmel} S.,  2014, \mn@doi [\apjl]
  {10.1088/2041-8205/784/1/L14}, \href
  {http://adsabs.harvard.edu/abs/2014ApJ...784L..14B} {784, L14}

\bibitem[\protect\citeauthoryear{{Bullock}, {Kolatt}, {Sigad}, {Somerville},
  {Kravtsov}, {Klypin}, {Primack}  \& {Dekel}}{{Bullock}
  et~al.}{2001}]{2001MNRAS.321..559B}
{Bullock} J.~S.,  {Kolatt} T.~S.,  {Sigad} Y.,  {Somerville} R.~S.,  {Kravtsov}
  A.~V.,  {Klypin} A.~A.,  {Primack} J.~R.,   {Dekel} A.,  2001, \mn@doi
  [\mnras] {10.1046/j.1365-8711.2001.04068.x}, \href
  {http://adsabs.harvard.edu/abs/2001MNRAS.321..559B} {321, 559}

\bibitem[\protect\citeauthoryear{{Catelan} \& {Theuns}}{{Catelan} \&
  {Theuns}}{1996}]{CatelanTheuns96}
{Catelan} P.,  {Theuns} T.,  1996, \mnras, \href
  {http://ukads.nottingham.ac.uk/abs/1996MNRAS.282..436C} {282, 436}

\bibitem[\protect\citeauthoryear{{Codis} et~al.,}{{Codis}
  et~al.}{2015}]{2015MNRAS.448.3391C}
{Codis} S.,  et~al., 2015, \mn@doi [\mnras] {10.1093/mnras/stv231}, \href
  {http://adsabs.harvard.edu/abs/2015MNRAS.448.3391C} {448, 3391}

\bibitem[\protect\citeauthoryear{{Correa}, {Wyithe}, {Schaye}  \&
  {Duffy}}{{Correa} et~al.}{2014}]{2014arXiv1409.5228C}
{Correa} C.~A.,  {Wyithe} J.~S.~B.,  {Schaye} J.,   {Duffy} A.~R.,  2014,
  preprint, \href {http://adsabs.harvard.edu/abs/2014arXiv1409.5228C} {}
  (\mn@eprint {arXiv} {1409.5228})

\bibitem[\protect\citeauthoryear{{Correa}, {Wyithe}, {Schaye}  \&
  {Duffy}}{{Correa} et~al.}{2015a}]{2015arXiv150104382C}
{Correa} C.~A.,  {Wyithe} J.~S.~B.,  {Schaye} J.,   {Duffy} A.~R.,  2015a,
  preprint, \href {http://adsabs.harvard.edu/abs/2015arXiv150104382C} {}
  (\mn@eprint {arXiv} {1501.04382})

\bibitem[\protect\citeauthoryear{{Correa}, {Wyithe}, {Schaye}  \&
  {Duffy}}{{Correa} et~al.}{2015b}]{2015arXiv150200391C}
{Correa} C.~A.,  {Wyithe} J.~S.~B.,  {Schaye} J.,   {Duffy} A.~R.,  2015b,
  preprint, \href {http://adsabs.harvard.edu/abs/2015arXiv150200391C} {}
  (\mn@eprint {arXiv} {1502.00391})

\bibitem[\protect\citeauthoryear{{Dayal}, {Libeskind}  \& {Dunlop}}{{Dayal}
  et~al.}{2013}]{CLUESDayal13}
{Dayal} P.,  {Libeskind} N.~I.,   {Dunlop} J.~S.,  2013, \mn@doi [\mnras]
  {10.1093/mnras/stt446}, \href
  {http://adsabs.harvard.edu/abs/2013MNRAS.431.3618D} {431, 3618}

\bibitem[\protect\citeauthoryear{{Doumler}, {Gottl{\"o}ber}, {Hoffman}  \&
  {Courtois}}{{Doumler} et~al.}{2013}]{Doumler13}
{Doumler} T.,  {Gottl{\"o}ber} S.,  {Hoffman} Y.,   {Courtois} H.,  2013,
  \mn@doi [\mnras] {10.1093/mnras/sts614}, \href
  {http://adsabs.harvard.edu/abs/2013MNRAS.430..912D} {430, 912}

\bibitem[\protect\citeauthoryear{{Dubois}, {Pichon}, {Haehnelt}, {Kimm},
  {Slyz}, {Devriendt}  \& {Pogosyan}}{{Dubois} et~al.}{2012}]{Dubois12}
{Dubois} Y.,  {Pichon} C.,  {Haehnelt} M.,  {Kimm} T.,  {Slyz} A.,  {Devriendt}
  J.,   {Pogosyan} D.,  2012, \mn@doi [\mnras]
  {10.1111/j.1365-2966.2012.21160.x}, \href
  {http://adsabs.harvard.edu/abs/2012MNRAS.423.3616D} {423, 3616}

\bibitem[\protect\citeauthoryear{{Duffy}, {Schaye}, {Kay}  \& {Dalla
  Vecchia}}{{Duffy} et~al.}{2008}]{2008MNRAS.390L..64D}
{Duffy} A.~R.,  {Schaye} J.,  {Kay} S.~T.,   {Dalla Vecchia} C.,  2008, \mn@doi
  [\mnras] {10.1111/j.1745-3933.2008.00537.x}, \href
  {http://adsabs.harvard.edu/abs/2008MNRAS.390L..64D} {390, L64}

\bibitem[\protect\citeauthoryear{{Dunkley} et~al.,}{{Dunkley}
  et~al.}{2009}]{2009ApJS..180..306D}
{Dunkley} J.,  et~al., 2009, \mn@doi [\apjs] {10.1088/0067-0049/180/2/306},
  \href {http://adsabs.harvard.edu/abs/2009ApJS..180..306D} {180, 306}

\bibitem[\protect\citeauthoryear{{Forero-Romero}, {Hoffman}, {Yepes},
  {Gottl{\"o}ber}, {Piontek}, {Klypin}  \& {Steinmetz}}{{Forero-Romero}
  et~al.}{2011}]{CLUESForeroRomero11}
{Forero-Romero} J.~E.,  {Hoffman} Y.,  {Yepes} G.,  {Gottl{\"o}ber} S.,
  {Piontek} R.,  {Klypin} A.,   {Steinmetz} M.,  2011, \mn@doi [\mnras]
  {10.1111/j.1365-2966.2011.19358.x}, \href
  {http://adsabs.harvard.edu/abs/2011MNRAS.417.1434F} {417, 1434}

\bibitem[\protect\citeauthoryear{{Frenk} et~al.,}{{Frenk}
  et~al.}{1999}]{Frenk99}
{Frenk} C.~S.,  et~al., 1999, \mn@doi [\apj] {10.1086/307908}, \href
  {http://ukads.nottingham.ac.uk/abs/1999ApJ...525..554F} {525, 554}

\bibitem[\protect\citeauthoryear{{Genel} et~al.,}{{Genel}
  et~al.}{2014}]{2014MNRAS.445..175G}
{Genel} S.,  et~al., 2014, \mn@doi [\mnras] {10.1093/mnras/stu1654}, \href
  {http://adsabs.harvard.edu/abs/2014MNRAS.445..175G} {445, 175}

\bibitem[\protect\citeauthoryear{{Governato}, {Willman}, {Mayer}, {Brooks},
  {Stinson}, {Valenzuela}, {Wadsley}  \& {Quinn}}{{Governato}
  et~al.}{2007}]{2007MNRAS.374.1479G}
{Governato} F.,  {Willman} B.,  {Mayer} L.,  {Brooks} A.,  {Stinson} G.,
  {Valenzuela} O.,  {Wadsley} J.,   {Quinn} T.,  2007, \mn@doi [\mnras]
  {10.1111/j.1365-2966.2006.11266.x}, \href
  {http://adsabs.harvard.edu/abs/2007MNRAS.374.1479G} {374, 1479}

\bibitem[\protect\citeauthoryear{{Guedes}, {Callegari}, {Madau}  \&
  {Mayer}}{{Guedes} et~al.}{2011}]{2011ApJ...742...76G}
{Guedes} J.,  {Callegari} S.,  {Madau} P.,   {Mayer} L.,  2011, \mn@doi [\apj]
  {10.1088/0004-637X/742/2/76}, \href
  {http://adsabs.harvard.edu/abs/2011ApJ...742...76G} {742, 76}

\bibitem[\protect\citeauthoryear{{Hahn} \& {Abel}}{{Hahn} \&
  {Abel}}{2011}]{2011MNRAS.415.2101H}
{Hahn} O.,  {Abel} T.,  2011, \mn@doi [\mnras]
  {10.1111/j.1365-2966.2011.18820.x}, \href
  {http://adsabs.harvard.edu/abs/2011MNRAS.415.2101H} {415, 2101}

\bibitem[\protect\citeauthoryear{{He{\ss}}, {Kitaura}  \&
  {Gottl{\"o}ber}}{{He{\ss}} et~al.}{2013}]{Hess13}
{He{\ss}} S.,  {Kitaura} F.-S.,   {Gottl{\"o}ber} S.,  2013, \mn@doi [\mnras]
  {10.1093/mnras/stt1428}, \href
  {http://adsabs.harvard.edu/abs/2013MNRAS.435.2065H} {435, 2065}

\bibitem[\protect\citeauthoryear{{Hoffman} \& {Ribak}}{{Hoffman} \&
  {Ribak}}{1991}]{HR91}
{Hoffman} Y.,  {Ribak} E.,  1991, \mn@doi [\apjl] {10.1086/186160}, \href
  {http://adsabs.harvard.edu/abs/1991ApJ...380L...5H} {380, L5}

\bibitem[\protect\citeauthoryear{{Hoffman}, {Romano-D{\'{\i}}az}, {Shlosman}
  \& {Heller}}{{Hoffman} et~al.}{2007}]{Hoffman07}
{Hoffman} Y.,  {Romano-D{\'{\i}}az} E.,  {Shlosman} I.,   {Heller} C.,  2007,
  \mn@doi [\apj] {10.1086/523695}, \href
  {http://ukads.nottingham.ac.uk/abs/2007ApJ...671.1108H} {671, 1108}

\bibitem[\protect\citeauthoryear{{Hopkins}, {Cox}, {Younger}  \&
  {Hernquist}}{{Hopkins} et~al.}{2009}]{2009ApJ...691.1168H}
{Hopkins} P.~F.,  {Cox} T.~J.,  {Younger} J.~D.,   {Hernquist} L.,  2009,
  \mn@doi [\apj] {10.1088/0004-637X/691/2/1168}, \href
  {http://adsabs.harvard.edu/abs/2009ApJ...691.1168H} {691, 1168}

\bibitem[\protect\citeauthoryear{{Hopkins}, {Cox}, {Hernquist}, {Narayanan},
  {Hayward}  \& {Murray}}{{Hopkins} et~al.}{2013}]{2013MNRAS.430.1901H}
{Hopkins} P.~F.,  {Cox} T.~J.,  {Hernquist} L.,  {Narayanan} D.,  {Hayward}
  C.~C.,   {Murray} N.,  2013, \mn@doi [\mnras] {10.1093/mnras/stt017}, \href
  {http://adsabs.harvard.edu/abs/2013MNRAS.430.1901H} {430, 1901}

\bibitem[\protect\citeauthoryear{{Iliev}, {Moore}, {Gottl{\"o}ber}, {Yepes},
  {Hoffman}  \& {Mellema}}{{Iliev} et~al.}{2011}]{Iliev11}
{Iliev} I.~T.,  {Moore} B.,  {Gottl{\"o}ber} S.,  {Yepes} G.,  {Hoffman} Y.,
  {Mellema} G.,  2011, \mn@doi [\mnras] {10.1111/j.1365-2966.2011.18292.x},
  \href {http://adsabs.harvard.edu/abs/2011MNRAS.413.2093I} {413, 2093}

\bibitem[\protect\citeauthoryear{{Jasche} \& {Wandelt}}{{Jasche} \&
  {Wandelt}}{2013}]{2013MNRAS.432..894J}
{Jasche} J.,  {Wandelt} B.~D.,  2013, \mn@doi [\mnras] {10.1093/mnras/stt449},
  \href {http://adsabs.harvard.edu/abs/2013MNRAS.432..894J} {432, 894}

\bibitem[\protect\citeauthoryear{{Jeeson-Daniel}, {Dalla Vecchia}, {Haas}  \&
  {Schaye}}{{Jeeson-Daniel} et~al.}{2011}]{2011MNRAS.415L..69J}
{Jeeson-Daniel} A.,  {Dalla Vecchia} C.,  {Haas} M.~R.,   {Schaye} J.,  2011,
  \mn@doi [\mnras] {10.1111/j.1745-3933.2011.01081.x}, \href
  {http://adsabs.harvard.edu/abs/2011MNRAS.415L..69J} {415, L69}

\bibitem[\protect\citeauthoryear{{Kitaura}}{{Kitaura}}{2013}]{Kitaura13}
{Kitaura} F.-S.,  2013, \mn@doi [\mnras] {10.1093/mnrasl/sls029}, \href
  {http://adsabs.harvard.edu/abs/2013MNRAS.429L..84K} {429, L84}

\bibitem[\protect\citeauthoryear{{Klimentowski}, {{\L}okas}, {Knebe},
  {Gottl{\"o}ber}, {Martinez-Vaquero}, {Yepes}  \& {Hoffman}}{{Klimentowski}
  et~al.}{2010}]{Klimentowski2010}
{Klimentowski} J.,  {{\L}okas} E.~L.,  {Knebe} A.,  {Gottl{\"o}ber} S.,
  {Martinez-Vaquero} L.~A.,  {Yepes} G.,   {Hoffman} Y.,  2010, \mn@doi
  [\mnras] {10.1111/j.1365-2966.2009.16024.x}, \href
  {http://ukads.nottingham.ac.uk/abs/2010MNRAS.402.1899K} {402, 1899}

\bibitem[\protect\citeauthoryear{{Klypin}, {Hoffman}, {Kravtsov}  \&
  {Gottl{\"o}ber}}{{Klypin} et~al.}{2003}]{Klypin2003}
{Klypin} A.,  {Hoffman} Y.,  {Kravtsov} A.~V.,   {Gottl{\"o}ber} S.,  2003,
  \mn@doi [\apj] {10.1086/377574}, \href
  {http://ukads.nottingham.ac.uk/abs/2003ApJ...596...19K} {596, 19}

\bibitem[\protect\citeauthoryear{{Klypin}, {Yepes}, {Gottlober}, {Prada}  \&
  {Hess}}{{Klypin} et~al.}{2014}]{2014arXiv1411.4001K}
{Klypin} A.,  {Yepes} G.,  {Gottlober} S.,  {Prada} F.,   {Hess} S.,  2014,
  preprint, \href {http://adsabs.harvard.edu/abs/2014arXiv1411.4001K} {}
  (\mn@eprint {arXiv} {1411.4001})

\bibitem[\protect\citeauthoryear{{Kormendy}}{{Kormendy}}{2015}]{2015arXiv150403330K}
{Kormendy} J.,  2015, in Laurikainen E.,  ed., , Galactic Bulges.
Springer, New York (\mn@eprint {arXiv} {1504.03330})

\bibitem[\protect\citeauthoryear{{Kravtsov}, {Klypin}  \& {Hoffman}}{{Kravtsov}
  et~al.}{2002}]{Kravtsov2002}
{Kravtsov} A.~V.,  {Klypin} A.,   {Hoffman} Y.,  2002, \mn@doi [\apj]
  {10.1086/340046}, \href
  {http://ukads.nottingham.ac.uk/abs/2002ApJ...571..563K} {571, 563}

\bibitem[\protect\citeauthoryear{{Libeskind}, {Yepes}, {Knebe},
  {Gottl{\"o}ber}, {Hoffman}  \& {Knollmann}}{{Libeskind}
  et~al.}{2010}]{Libeskind2010}
{Libeskind} N.~I.,  {Yepes} G.,  {Knebe} A.,  {Gottl{\"o}ber} S.,  {Hoffman}
  Y.,   {Knollmann} S.~R.,  2010, \mn@doi [\mnras]
  {10.1111/j.1365-2966.2009.15766.x}, \href
  {http://ukads.nottingham.ac.uk/abs/2010MNRAS.401.1889L} {401, 1889}

\bibitem[\protect\citeauthoryear{{Ludlow} et~al.,}{{Ludlow}
  et~al.}{2013}]{2013MNRAS.432.1103L}
{Ludlow} A.~D.,  et~al., 2013, \mn@doi [\mnras] {10.1093/mnras/stt526}, \href
  {http://adsabs.harvard.edu/abs/2013MNRAS.432.1103L} {432, 1103}

\bibitem[\protect\citeauthoryear{{Ludlow}, {Navarro}, {Angulo},
  {Boylan-Kolchin}, {Springel}, {Frenk}  \& {White}}{{Ludlow}
  et~al.}{2014}]{2014MNRAS.441..378L}
{Ludlow} A.~D.,  {Navarro} J.~F.,  {Angulo} R.~E.,  {Boylan-Kolchin} M.,
  {Springel} V.,  {Frenk} C.,   {White} S.~D.~M.,  2014, \mn@doi [\mnras]
  {10.1093/mnras/stu483}, \href
  {http://adsabs.harvard.edu/abs/2014MNRAS.441..378L} {441, 378}

\bibitem[\protect\citeauthoryear{{Macci{\`o}}, {Dutton}, {van den Bosch},
  {Moore}, {Potter}  \& {Stadel}}{{Macci{\`o}}
  et~al.}{2007}]{2007MNRAS.378...55M}
{Macci{\`o}} A.~V.,  {Dutton} A.~A.,  {van den Bosch} F.~C.,  {Moore} B.,
  {Potter} D.,   {Stadel} J.,  2007, \mn@doi [\mnras]
  {10.1111/j.1365-2966.2007.11720.x}, \href
  {http://adsabs.harvard.edu/abs/2007MNRAS.378...55M} {378, 55}

\bibitem[\protect\citeauthoryear{{Macci{\`o}}, {Dutton}  \& {van den
  Bosch}}{{Macci{\`o}} et~al.}{2008}]{2008MNRAS.391.1940M}
{Macci{\`o}} A.~V.,  {Dutton} A.~A.,   {van den Bosch} F.~C.,  2008, \mn@doi
  [\mnras] {10.1111/j.1365-2966.2008.14029.x}, \href
  {http://adsabs.harvard.edu/abs/2008MNRAS.391.1940M} {391, 1940}

\bibitem[\protect\citeauthoryear{{Mathis}, {Lemson}, {Springel}, {Kauffmann},
  {White}, {Eldar}  \& {Dekel}}{{Mathis} et~al.}{2002}]{Mathis02}
{Mathis} H.,  {Lemson} G.,  {Springel} V.,  {Kauffmann} G.,  {White} S.~D.~M.,
  {Eldar} A.,   {Dekel} A.,  2002, \mn@doi [\mnras]
  {10.1046/j.1365-8711.2002.05447.x}, \href
  {http://ukads.nottingham.ac.uk/abs/2002MNRAS.333..739M} {333, 739}

\bibitem[\protect\citeauthoryear{{McBride}, {Fakhouri}  \& {Ma}}{{McBride}
  et~al.}{2009}]{2009MNRAS.398.1858M}
{McBride} J.,  {Fakhouri} O.,   {Ma} C.-P.,  2009, \mn@doi [\mnras]
  {10.1111/j.1365-2966.2009.15329.x}, \href
  {http://adsabs.harvard.edu/abs/2009MNRAS.398.1858M} {398, 1858}

\bibitem[\protect\citeauthoryear{{Murray}, {Power}  \& {Robotham}}{{Murray}
  et~al.}{2013}]{2013A&C.....3...23M}
{Murray} S.~G.,  {Power} C.,   {Robotham} A.~S.~G.,  2013, \mn@doi [Astronomy
  and Computing] {10.1016/j.ascom.2013.11.001}, \href
  {http://adsabs.harvard.edu/abs/2013A%26C.....3...23M} {3, 23}

\bibitem[\protect\citeauthoryear{{Naab}, {Burkert}  \& {Hernquist}}{{Naab}
  et~al.}{1999}]{1999ApJ...523L.133N}
{Naab} T.,  {Burkert} A.,   {Hernquist} L.,  1999, \mn@doi [\apjl]
  {10.1086/312275}, \href {http://adsabs.harvard.edu/abs/1999ApJ...523L.133N}
  {523, L133}

\bibitem[\protect\citeauthoryear{{Navarro}, {Frenk}  \& {White}}{{Navarro}
  et~al.}{1997}]{1997ApJ...490..493N}
{Navarro} J.~F.,  {Frenk} C.~S.,   {White} S.~D.~M.,  1997, \mn@doi [\apj]
  {10.1086/304888}, \href {http://adsabs.harvard.edu/abs/1997ApJ...490..493N}
  {490, 493}

\bibitem[\protect\citeauthoryear{{Neto} et~al.,}{{Neto}
  et~al.}{2007}]{2007MNRAS.381.1450N}
{Neto} A.~F.,  et~al., 2007, \mn@doi [\mnras]
  {10.1111/j.1365-2966.2007.12381.x}, \href
  {http://adsabs.harvard.edu/abs/2007MNRAS.381.1450N} {381, 1450}

\bibitem[\protect\citeauthoryear{{Nuza}, {Parisi}, {Scannapieco}, {Richter},
  {Gottl{\"o}ber}  \& {Steinmetz}}{{Nuza} et~al.}{2014}]{CLUESNuza14}
{Nuza} S.~E.,  {Parisi} F.,  {Scannapieco} C.,  {Richter} P.,  {Gottl{\"o}ber}
  S.,   {Steinmetz} M.,  2014, \mn@doi [\mnras] {10.1093/mnras/stu643}, \href
  {http://adsabs.harvard.edu/abs/2014MNRAS.441.2593N} {441, 2593}

\bibitem[\protect\citeauthoryear{{Papovich} et~al.,}{{Papovich}
  et~al.}{2015}]{2015ApJ...803...26P}
{Papovich} C.,  et~al., 2015, \mn@doi [\apj] {10.1088/0004-637X/803/1/26},
  \href {http://adsabs.harvard.edu/abs/2015ApJ...803...26P} {803, 26}

\bibitem[\protect\citeauthoryear{{Pontzen} \& {Governato}}{{Pontzen} \&
  {Governato}}{2014}]{2014Natur.506..171P}
{Pontzen} A.,  {Governato} F.,  2014, \mn@doi [\nat] {10.1038/nature12953},
  \href {http://adsabs.harvard.edu/abs/2014Natur.506..171P} {506, 171}

\bibitem[\protect\citeauthoryear{{Pontzen}, {Ro{\v s}kar}, {Stinson}  \&
  {Woods}}{{Pontzen} et~al.}{2013}]{2013ascl.soft05002P}
{Pontzen} A.,  {Ro{\v s}kar} R.,  {Stinson} G.,   {Woods} R.,  2013, {pynbody:
  N-Body/SPH analysis for python}, Astrophysics Source Code Library (\mn@eprint
  {ascl} {1305.002})

\bibitem[\protect\citeauthoryear{{Porciani}, {Dekel}  \& {Hoffman}}{{Porciani}
  et~al.}{2002a}]{Porciani02a}
{Porciani} C.,  {Dekel} A.,   {Hoffman} Y.,  2002a, \mn@doi [\mnras]
  {10.1046/j.1365-8711.2002.05305.x}, \href
  {http://ukads.nottingham.ac.uk/abs/2002MNRAS.332..325P} {332, 325}

\bibitem[\protect\citeauthoryear{{Porciani}, {Dekel}  \& {Hoffman}}{{Porciani}
  et~al.}{2002b}]{Porciani02b}
{Porciani} C.,  {Dekel} A.,   {Hoffman} Y.,  2002b, \mn@doi [\mnras]
  {10.1046/j.1365-8711.2002.05306.x}, \href
  {http://adsabs.harvard.edu/abs/2002MNRAS.332..339P} {332, 339}

\bibitem[\protect\citeauthoryear{{Power}, {Navarro}, {Jenkins}, {Frenk},
  {White}, {Springel}, {Stadel}  \& {Quinn}}{{Power} et~al.}{2003}]{power}
{Power} C.,  {Navarro} J.~F.,  {Jenkins} A.,  {Frenk} C.~S.,  {White} S.~D.~M.,
   {Springel} V.,  {Stadel} J.,   {Quinn} T.,  2003, \mn@doi [\mnras]
  {10.1046/j.1365-8711.2003.05925.x}, \href
  {http://adsabs.harvard.edu/abs/2003MNRAS.338...14P} {338, 14}

\bibitem[\protect\citeauthoryear{{Prada}, {Klypin}, {Cuesta}, {Betancort-Rijo}
  \& {Primack}}{{Prada} et~al.}{2012}]{2012MNRAS.423.3018P}
{Prada} F.,  {Klypin} A.~A.,  {Cuesta} A.~J.,  {Betancort-Rijo} J.~E.,
  {Primack} J.,  2012, \mn@doi [\mnras] {10.1111/j.1365-2966.2012.21007.x},
  \href {http://adsabs.harvard.edu/abs/2012MNRAS.423.3018P} {423, 3018}

\bibitem[\protect\citeauthoryear{{Press} \& {Schechter}}{{Press} \&
  {Schechter}}{1974}]{1974ApJ...187..425P}
{Press} W.~H.,  {Schechter} P.,  1974, \mn@doi [\apj] {10.1086/152650}, \href
  {http://adsabs.harvard.edu/abs/1974ApJ...187..425P} {187, 425}

\bibitem[\protect\citeauthoryear{{Prunet}, {Pichon}, {Aubert}, {Pogosyan},
  {Teyssier}  \& {Gottloeber}}{{Prunet} et~al.}{2008}]{Prunet08MPGRAFIC}
{Prunet} S.,  {Pichon} C.,  {Aubert} D.,  {Pogosyan} D.,  {Teyssier} R.,
  {Gottloeber} S.,  2008, \mn@doi [\apjs] {10.1086/590370}, \href
  {http://adsabs.harvard.edu/abs/2008ApJS..178..179P} {178, 179}

\bibitem[\protect\citeauthoryear{{Ragone-Figueroa}, {Plionis}, {Merch{\'a}n},
  {Gottl{\"o}ber}  \& {Yepes}}{{Ragone-Figueroa}
  et~al.}{2010}]{2010MNRAS.407..581R}
{Ragone-Figueroa} C.,  {Plionis} M.,  {Merch{\'a}n} M.,  {Gottl{\"o}ber} S.,
  {Yepes} G.,  2010, \mn@doi [\mnras] {10.1111/j.1365-2966.2010.16935.x}, \href
  {http://adsabs.harvard.edu/abs/2010MNRAS.407..581R} {407, 581}

\bibitem[\protect\citeauthoryear{{Reed}, {Governato}, {Verde}, {Gardner},
  {Quinn}, {Stadel}, {Merritt}  \& {Lake}}{{Reed}
  et~al.}{2005}]{2005MNRAS.357...82R}
{Reed} D.,  {Governato} F.,  {Verde} L.,  {Gardner} J.,  {Quinn} T.,  {Stadel}
  J.,  {Merritt} D.,   {Lake} G.,  2005, \mn@doi [\mnras]
  {10.1111/j.1365-2966.2005.08612.x}, \href
  {http://adsabs.harvard.edu/abs/2005MNRAS.357...82R} {357, 82}

\bibitem[\protect\citeauthoryear{{Robertson}, {Bullock}, {Cox}, {Di Matteo},
  {Hernquist}, {Springel}  \& {Yoshida}}{{Robertson}
  et~al.}{2006}]{2006ApJ...645..986R}
{Robertson} B.,  {Bullock} J.~S.,  {Cox} T.~J.,  {Di Matteo} T.,  {Hernquist}
  L.,  {Springel} V.,   {Yoshida} N.,  2006, \mn@doi [\apj] {10.1086/504412},
  \href {http://adsabs.harvard.edu/abs/2006ApJ...645..986R} {645, 986}

\bibitem[\protect\citeauthoryear{{Romano-D{\'{\i}}az}, {Faltenbacher}, {Jones},
  {Heller}, {Hoffman}  \& {Shlosman}}{{Romano-D{\'{\i}}az}
  et~al.}{2006}]{RomanoDiaz06}
{Romano-D{\'{\i}}az} E.,  {Faltenbacher} A.,  {Jones} D.,  {Heller} C.,
  {Hoffman} Y.,   {Shlosman} I.,  2006, \mn@doi [\apjl] {10.1086/500645}, \href
  {http://ukads.nottingham.ac.uk/abs/2006ApJ...637L..93R} {637, L93}

\bibitem[\protect\citeauthoryear{{Romano-D{\'{\i}}az}, {Hoffman}, {Heller},
  {Faltenbacher}, {Jones}  \& {Shlosman}}{{Romano-D{\'{\i}}az}
  et~al.}{2007}]{RomanoDiaz07}
{Romano-D{\'{\i}}az} E.,  {Hoffman} Y.,  {Heller} C.,  {Faltenbacher} A.,
  {Jones} D.,   {Shlosman} I.,  2007, \mn@doi [\apj] {10.1086/509798}, \href
  {http://ukads.nottingham.ac.uk/abs/2007ApJ...657...56R} {657, 56}

\bibitem[\protect\citeauthoryear{{Romano-D{\'{\i}}az}, {Shlosman}, {Trenti}  \&
  {Hoffman}}{{Romano-D{\'{\i}}az} et~al.}{2011a}]{RomanoDiaz2011a}
{Romano-D{\'{\i}}az} E.,  {Shlosman} I.,  {Trenti} M.,   {Hoffman} Y.,  2011a,
  \mn@doi [\apj] {10.1088/0004-637X/736/1/66}, \href
  {http://adsabs.harvard.edu/abs/2011ApJ...736...66R} {736, 66}

\bibitem[\protect\citeauthoryear{{Romano-D{\'{\i}}az}, {Choi}, {Shlosman}  \&
  {Trenti}}{{Romano-D{\'{\i}}az} et~al.}{2011b}]{RomanoDiaz2011b}
{Romano-D{\'{\i}}az} E.,  {Choi} J.-H.,  {Shlosman} I.,   {Trenti} M.,  2011b,
  \mn@doi [\apjl] {10.1088/2041-8205/738/2/L19}, \href
  {http://adsabs.harvard.edu/abs/2011ApJ...738L..19R} {738, L19}

\bibitem[\protect\citeauthoryear{{Romano-D{\'{\i}}az}, {Shlosman}, {Choi}  \&
  {Sadoun}}{{Romano-D{\'{\i}}az} et~al.}{2014}]{RomanoDiaz2014}
{Romano-D{\'{\i}}az} E.,  {Shlosman} I.,  {Choi} J.-H.,   {Sadoun} R.,  2014,
  \mn@doi [\apjl] {10.1088/2041-8205/790/2/L32}, \href
  {http://adsabs.harvard.edu/abs/2014ApJ...790L..32R} {790, L32}

\bibitem[\protect\citeauthoryear{{Schaye} et~al.,}{{Schaye}
  et~al.}{2015}]{2015MNRAS.446..521S}
{Schaye} J.,  et~al., 2015, \mn@doi [\mnras] {10.1093/mnras/stu2058}, \href
  {http://adsabs.harvard.edu/abs/2015MNRAS.446..521S} {446, 521}

\bibitem[\protect\citeauthoryear{Sherman \& Morrison}{Sherman \&
  Morrison}{1950}]{sherman1950}
Sherman J.,  Morrison W.~J.,  1950, \mn@doi [Ann. Math. Statist.]
  {10.1214/aoms/1177729893}, 21, 124

\bibitem[\protect\citeauthoryear{{Skibba} \& {Macci{\`o}}}{{Skibba} \&
  {Macci{\`o}}}{2011}]{2011MNRAS.416.2388S}
{Skibba} R.~A.,  {Macci{\`o}} A.~V.,  2011, \mn@doi [\mnras]
  {10.1111/j.1365-2966.2011.19218.x}, \href
  {http://adsabs.harvard.edu/abs/2011MNRAS.416.2388S} {416, 2388}

\bibitem[\protect\citeauthoryear{{Sorce}, {Courtois}, {Gottl{\"o}ber},
  {Hoffman}  \& {Tully}}{{Sorce} et~al.}{2014}]{Sorce14}
{Sorce} J.~G.,  {Courtois} H.~M.,  {Gottl{\"o}ber} S.,  {Hoffman} Y.,   {Tully}
  R.~B.,  2014, \mn@doi [\mnras] {10.1093/mnras/stt2153}, \href
  {http://adsabs.harvard.edu/abs/2014MNRAS.437.3586S} {437, 3586}

\bibitem[\protect\citeauthoryear{{Springel}}{{Springel}}{2005}]{Springel2005}
{Springel} V.,  2005, \mn@doi [\mnras] {10.1111/j.1365-2966.2005.09655.x},
  \href {http://adsabs.harvard.edu/abs/2005MNRAS.364.1105S} {364, 1105}

\bibitem[\protect\citeauthoryear{{Springel}, {White}, {Tormen}  \&
  {Kauffmann}}{{Springel} et~al.}{2001}]{2001MNRAS.328..726S}
{Springel} V.,  {White} S.~D.~M.,  {Tormen} G.,   {Kauffmann} G.,  2001,
  \mn@doi [\mnras] {10.1046/j.1365-8711.2001.04912.x}, \href
  {http://adsabs.harvard.edu/abs/2001MNRAS.328..726S} {328, 726}

\bibitem[\protect\citeauthoryear{{Springel} et~al.,}{{Springel}
  et~al.}{2008}]{2008MNRAS.391.1685S}
{Springel} V.,  et~al., 2008, \mn@doi [\mnras]
  {10.1111/j.1365-2966.2008.14066.x}, \href
  {http://adsabs.harvard.edu/abs/2008MNRAS.391.1685S} {391, 1685}

\bibitem[\protect\citeauthoryear{{Tasitsiomi}, {Kravtsov}, {Gottl{\"o}ber}  \&
  {Klypin}}{{Tasitsiomi} et~al.}{2004}]{2004ApJ...607..125T}
{Tasitsiomi} A.,  {Kravtsov} A.~V.,  {Gottl{\"o}ber} S.,   {Klypin} A.~A.,
  2004, \mn@doi [\apj] {10.1086/383219}, \href
  {http://adsabs.harvard.edu/abs/2004ApJ...607..125T} {607, 125}

\bibitem[\protect\citeauthoryear{{Tinker}, {Kravtsov}, {Klypin}, {Abazajian},
  {Warren}, {Yepes}, {Gottl{\"o}ber}  \& {Holz}}{{Tinker}
  et~al.}{2008}]{2008ApJ...688..709T}
{Tinker} J.,  {Kravtsov} A.~V.,  {Klypin} A.,  {Abazajian} K.,  {Warren} M.,
  {Yepes} G.,  {Gottl{\"o}ber} S.,   {Holz} D.~E.,  2008, \mn@doi [\apj]
  {10.1086/591439}, \href {http://adsabs.harvard.edu/abs/2008ApJ...688..709T}
  {688, 709}

\bibitem[\protect\citeauthoryear{{Wechsler}, {Bullock}, {Primack}, {Kravtsov}
  \& {Dekel}}{{Wechsler} et~al.}{2002}]{2002ApJ...568...52W}
{Wechsler} R.~H.,  {Bullock} J.~S.,  {Primack} J.~R.,  {Kravtsov} A.~V.,
  {Dekel} A.,  2002, \mn@doi [\apj] {10.1086/338765}, \href
  {http://adsabs.harvard.edu/abs/2002ApJ...568...52W} {568, 52}

\bibitem[\protect\citeauthoryear{{White}}{{White}}{1984}]{1984ApJ...286...38W}
{White} S.~D.~M.,  1984, \mn@doi [\apj] {10.1086/162573}, \href
  {http://adsabs.harvard.edu/abs/1984ApJ...286...38W} {286, 38}

\bibitem[\protect\citeauthoryear{{Wong} \& {Taylor}}{{Wong} \&
  {Taylor}}{2012}]{2012ApJ...757..102W}
{Wong} A.~W.~C.,  {Taylor} J.~E.,  2012, \mn@doi [\apj]
  {10.1088/0004-637X/757/1/102}, \href
  {http://adsabs.harvard.edu/abs/2012ApJ...757..102W} {757, 102}

\bibitem[\protect\citeauthoryear{{Zaroubi}, {Hoffman}, {Fisher}  \&
  {Lahav}}{{Zaroubi} et~al.}{1995}]{1995ApJ...449..446Z}
{Zaroubi} S.,  {Hoffman} Y.,  {Fisher} K.~B.,   {Lahav} O.,  1995, \mn@doi
  [\apj] {10.1086/176070}, \href
  {http://adsabs.harvard.edu/abs/1995ApJ...449..446Z} {449, 446}

\bibitem[\protect\citeauthoryear{{Zavala}, {Jing}, {Faltenbacher}, {Yepes},
  {Hoffman}, {Gottl{\"o}ber}  \& {Catinella}}{{Zavala} et~al.}{2009}]{Zavala09}
{Zavala} J.,  {Jing} Y.~P.,  {Faltenbacher} A.,  {Yepes} G.,  {Hoffman} Y.,
  {Gottl{\"o}ber} S.,   {Catinella} B.,  2009, \mn@doi [\apj]
  {10.1088/0004-637X/700/2/1779}, \href
  {http://ukads.nottingham.ac.uk/abs/2009ApJ...700.1779Z} {700, 1779}

\bibitem[\protect\citeauthoryear{{Zhao}, {Mo}, {Jing}  \& {B{\"o}rner}}{{Zhao}
  et~al.}{2003}]{2003MNRAS.339...12Z}
{Zhao} D.~H.,  {Mo} H.~J.,  {Jing} Y.~P.,   {B{\"o}rner} G.,  2003, \mn@doi
  [\mnras] {10.1046/j.1365-8711.2003.06135.x}, \href
  {http://adsabs.harvard.edu/abs/2003MNRAS.339...12Z} {339, 12}

\bibitem[\protect\citeauthoryear{{Zhao}, {Jing}, {Mo}  \& {B{\"o}rner}}{{Zhao}
  et~al.}{2009}]{2009ApJ...707..354Z}
{Zhao} D.~H.,  {Jing} Y.~P.,  {Mo} H.~J.,   {B{\"o}rner} G.,  2009, \mn@doi
  [\apj] {10.1088/0004-637X/707/1/354}, \href
  {http://adsabs.harvard.edu/abs/2009ApJ...707..354Z} {707, 354}

\bibitem[\protect\citeauthoryear{{van Dokkum} et~al.,}{{van Dokkum}
  et~al.}{2013}]{2013ApJ...771L..35V}
{van Dokkum} P.~G.,  et~al., 2013, \mn@doi [\apjl]
  {10.1088/2041-8205/771/2/L35}, \href
  {http://adsabs.harvard.edu/abs/2013ApJ...771L..35V} {771, L35}

\bibitem[\protect\citeauthoryear{{van de Weygaert} \& {Bertschinger}}{{van de
  Weygaert} \& {Bertschinger}}{1996}]{1996MNRAS.281...84V}
{van de Weygaert} R.,  {Bertschinger} E.,  1996, \mnras, \href
  {http://adsabs.harvard.edu/abs/1996MNRAS.281...84V} {281, 84}

\bibitem[\protect\citeauthoryear{{van den Bosch}}{{van den
  Bosch}}{2002}]{2002MNRAS.331...98V}
{van den Bosch} F.~C.,  2002, \mn@doi [\mnras]
  {10.1046/j.1365-8711.2002.05171.x}, \href
  {http://adsabs.harvard.edu/abs/2002MNRAS.331...98V} {331, 98}

\makeatother
\end{thebibliography}

\appendix

\section{Technique}
\label{app-tech}
Here, we will give a detailed description of the derivation of the HR91 operator and our Eq.~\eqref{eq-dn-full}.

The values of the three-dimensional overdensity field $\vec{\delta} \equiv \delta(\vec{x})$ in the initial conditions of our cosmological simulations are distributed according to a multivariate Gaussian
\begin{equation}
p_0(\vec{\delta}) \propto \exp \left(-\frac{1}{2} (\vec{\delta} -
  \vec{\mu}_0)^{\dagger} \tens{C}_0^{-1} (\vec{\delta}-\vec{\mu}_0) \right)\textrm{,}
\end{equation}
with mean $\langle \vec{\delta} \rangle = \vec{\mu}_0$ and
$\tens{C}_0 = \langle (\vec{\delta}- \vec{\mu}_0 )^{\dagger} (\vec{\delta} - \vec{\mu}_0 ) \rangle $ (the covariance, or the power spectrum in Fourier space).
Cosmological initial conditions have zero mean, $\vec{\mu}_0 = 0$,
but we will consider the fully general case.

We will build the general procedure by induction. Suppose we have a
$p_{i-1}(\vec{\delta})$ that describes the probability distribution
function for $i-1$ constraints; we now want to add the $i$th
constraint, ensuring that $\vec{\alpha}_i^{\dagger} \vec{\delta}  = d_i$
for some constraint vector $\vec{\alpha}_i$ and constant $d_i$. To gain samples from the
constrained distribution one could sample from the original
distribution and reject all those trials which lie too far away from
$\left | \vec{\alpha}_i^{\dagger} \vec{\delta} -
  d_i\right|^2=0$. Mathematically this can be expressed by multiplying
the original probability distribution by a penalty function, \eg
\begin{equation}
p_i(\vec{\delta}) \propto \lim_{\beta \to \infty} p_{i-1}(\vec{\delta}) \exp
\left(-\frac{\beta}{2} \left|\vec{\alpha}_i^{\dagger} \vec{\delta} - d_i\right|^2 \right)\textrm{,}
\label{eq:limiting-p1}
\end{equation}
where the constant of proportionality renormalizes the probability
distribution function and is dependent on $\beta$. In the limit $\beta\to\infty$, the penalty function becomes a Dirac-Delta distribution and the constraint is satisfied exactly.

Under the assumption that $p_{i-1}$ is Gaussian, the new probability function is the product of two Gaussians, and so
remains Gaussian itself; consequently after imposing $i$ constraints
we must be able to write
\begin{equation}
p_i(\vec{\delta}) \propto \exp \left( -\frac{1}{2}(\vec{\delta}-\vec{\mu}_i)^{\dagger} \tens{C}_i^{-1}
  (\vec{\delta}-\vec{\mu}_i) \right)
\label{eq:assumed-p1}
\end{equation}
for some mean $\vec{\mu}_i$ and covariance $\tens{C}_i$ which we will now derive. By
multiplying out Eq.~\eqref{eq:limiting-p1} we obtain 
\begin{align}
p_i(\vec{\delta}) & \propto \lim_{\beta \to \infty} \exp \left[-\frac{1}{2} (\vec{\delta} -
  \vec{\mu}_i)^{\dagger} \left(\tens{C}_{i-1}^{-1} + \beta \vec{\alpha}_i
    \vec{\alpha}_i^{\dagger} \right) (\vec{\delta}-\vec{\mu}_i) \right. \nonumber  \\
  -& \left. \vec{\mu}_i^{\dagger} \left(\tens{C}_{i-1}^{-1} + \beta \alpha_i
    \alpha_i^{\dagger} \right) \vec{\delta} +
  \vec{\mu}_{i-1}^{\dagger}\tens{C}_{i-1}^{-1}\vec{\delta} + \beta d_i
  \vec{\alpha}_i^{\dagger} \vec{\delta}  \right],
 \label{eq:plong}
\end{align}
where we have already thrown away several terms which are zero-order in
$\vec{\delta}$ since they just change the normalization. By comparing terms in Eqs.~\eqref{eq:assumed-p1} and \eqref{eq:plong} we can first read off $\tens{C}_i^{-1}=\tens{C}_{i-1}^{-1} + \beta \vec{\alpha}_i \vec{\alpha}_i^{\dagger}$.
We will also need a normalization for the $\vec{\alpha}_i$, which conveniently can
be chosen\footnote{Unless $\vec{\alpha}_i$ is a null direction of
  $\tens{C_{i-1}}$, but then there would be zero probability of our
  constraint in the original distribution.} as
\begin{equation}
\vec{\alpha}_i^{\dagger} \tens{C}_{i-1} \vec{\alpha}_i = 1\textrm{.}
\label{eq:norm}
\end{equation}
Next, we apply the Sherman-Morrison formula \citep{sherman1950},
\begin{align}
(\tens{C}_{i-1}^{-1} + \beta \vec{\alpha}_1 \vec{\alpha}_1^{\dagger})^{-1} &=\tens{C}_{i-1} - \beta \frac{\tens{C}_{i-1} \vec{\alpha}_i \vec{\alpha}_i^{\dagger}
  \tens{C}_{i-1}}{1+\beta \vec{\alpha}_i^{\dagger} \tens{C}_{i-1} \vec{\alpha}_i} \nonumber \\
&\simeq \tens{C}_{i-1} \left[ 1-(1-\beta^{-1}) \vec{\alpha}_i
  \vec{\alpha}_i^{\dagger} \tens{C}_{i-1} \right] \textrm{,}
  \label{eq-SMF}
\end{align}
where we have used $\beta \gg 1$ and the normalization condition
\eqref{eq:norm} in the second step.

The terms in the second line of Eq.~\eqref{eq:plong} have to cancel exactly. Plugging Eq.~\eqref{eq-SMF} into this expression leads to
\begin{align}
\vec{\mu}_i =  & \lim_{\beta \to \infty} \vec{\mu}_{i-1} - (1-\beta^{-1}) \tens{C}_{i-1} \vec{\alpha}_i \vec{\alpha}_i^{\dagger} \vec{\mu}_{i-1} + d_i \vec{\alpha}_i \tens{C}_{i-1} \vec{\alpha}_i^{\dagger} \nonumber \\
& = \vec{\mu}_{i-1} + \tens{C}_{i-1} \vec{\alpha}_i \left( d_i- \vec{\alpha}_i^{\dagger} \vec{\mu}_{i-1} \right),
\label{eq:app-mu}
\end{align}
and finally taking the limit in Eq.~\eqref{eq-SMF} yields
\begin{equation}
\tens{C}_i = \tens{C}_{i-1} - \tens{C}_{i-1}\vec{\alpha}_i\vec{\alpha}_i^{\dagger} \tens{C}_{i-1}.
\label{eq:app-CN}
\end{equation}
This result allows us to apply as many constraints
as desired analytically -- by looping over the constraints and updating
the covariance matrix and mean at each step, then drawing a constrained realization -- but this would be computationally expensive. Instead, the constrained realization can be constructed from the unconstrained field using a projection operator, which we will now derive.

For notational simplicity, in addition to
normalizing the constraints, it is also helpful to make them
orthogonal (\eg through a Gram-Schmidt procedure) in the sense
that $\vec{\alpha}_i^{\dagger} \tens{C}_0 \vec{\alpha}_j = 0$ for $i \ne
j$. One can then verify by substitution (see Sec.~\ref{app-ssec:comments}) that the constrained
field has mean
\begin{equation}
\vec{\mu}_n = \vec{\mu}_0 + \sum_{i=1}^n \tens{C}_0 \vec{\alpha}_i \left( d_i-
\vec{\alpha}_i^{\dagger} \vec{\mu}_0\right)
\label{eq-app-mean} 
\end{equation}
and covariance
\begin{equation}
\tens{C}_n = \tens{C}_0 - \sum_{i=1}^n
\tens{C}_0\vec{\alpha}_i\vec{\alpha}_i^{\dagger} \tens{C}_0 
\label{app-eq-cov}
\end{equation}
for orthonormalized $\left\{\vec{\alpha}_i\right\}$.

Efficiently drawing from the distribution implied by the above mean
and covariance is made possible by any operator $\tens{P}_n$ that
takes a realization from the unconstrained field $\vec{\delta}_0$ and
forms a new realization under $n$ constraints via the ansatz
\begin{equation}
\vec{\delta}_n = \tens{P}_n \left( \vec{\delta}_0 - \vec{\mu}_0
\right) + \vec{\mu}_n\textrm{,}
\label{app-eq-const}
\end{equation}
where to gain the correct covariance $\tens{C}_n = \langle (\vec{\delta}_n - \vec{\mu}_n) (\vec{\delta}_n - \vec{\mu}_n)^{\dagger} \rangle$ one must demand
\begin{equation}
\tens{P}_n \tens{C}_0 \tens{P}_n^{\dagger} = \tens{C}_n\textrm{.}
\label{app-eq:proj-requirement}
\end{equation}
There are an infinity of operators $\tens{P}_n$ with this property: Given any specific $\tens{P}_n$ one can form $\tens{P}_n' = \tens{U} \tens{P}_n$ where \mbox{$\tens{U}^{\dagger} \tens{C}_0 \tens{U} =
 \mathbb{1}$}, and the new $\tens{P}_n'$ satisfies the required identity
\eqref{app-eq:proj-requirement}. To obtain the unique HR91 operator, we additionally require $\tens{P}_n$ to make \emph{minimal} changes to the field. This implies $\tens{P}_n \vec{\delta}_n =\vec{\delta}_n$ --- in other words,
that no changes are made if the field already satisfies the constraints.
Using Eq.~\eqref{app-eq-const}, it immediately follows that $\tens{P}_n^2 = \tens{P}_n$ and
$\tens{P}_n \vec{\mu}_n = \vec{\mu}_n$. The first of these conditions
implies that all eigenvalues of $\tens{P}_n$ are either $1$ or $0$.

One can verify by substitution that all these requirements are satisfied by
\begin{equation}
\tens{P}_n = \mathbb{1} - \sum_{i=1}^n \tens{C}_0 \vec{\alpha}_i
\vec{\alpha}_i^{\dagger} \textrm{ (for orthonormalized
  $\left\{\vec{\alpha}_i\right\}$), }
\label{eq-app-HR91}
\end{equation}
Note that the HR91 form given in their Eqs. (2) -- (4) builds the
orthonormalization procedure into the projection operator (appearing
as $\xi_{ij}^{-1}$ in their notation). However, as stated above we
found it notationally simpler to
pre-condition the constraints into orthonormal form using the
Gram-Schmidt procedure. Both formulations are mathematically equivalent (see Appendix \ref{app-B}).

Inserting Eqs.~\eqref{eq-app-HR91} and \eqref{eq-app-mean} into \eqref{app-eq-const} then leads to the final expression
\begin{equation}
\vec{\delta}_n = \vec{\delta}_0 +\sum_{i=1}^n \C_0 \vec{\alpha}_i \left(d_i- d_{i0}\right),
\label{eq-app-final}
\end{equation}
where $d_{i0} = \vec{\alpha}_i^{\dagger} \vec{\delta}_0$.

In practice, most of the necessary calculations are performed in Fourier-space, because there $\tens{C}_0$ is the $\Lambda$CDM power spectrum which is diagonal. Any constraint vector $\vec{\alpha}_i$ and density field $\vec{\delta}$ can be easily converted using numerical Fast Fourier transformations.

Note that the algorithm in its current form only takes into account the contribution from the power spectrum. If one wanted to generate constrained initial conditions based on an observational dataset, the associated uncertainties would introduce extra contributions in the new covariance matrix \cite[\eg][]{1995ApJ...449..446Z, 1996MNRAS.281...84V}, which is not included in the current implementation.

\subsection{Comments on normalisation} 
\label{app-ssec:comments}
Throughout this paper, we use the same notation for the normalised and unnormalised constraints (expressed by $\vec{\alpha}_i^{\dagger} \vec{\delta} =d_i$). In practice, these quantities are affected by the normalisation condition in the following way: if $\vec{\alpha}^{\dagger}_i \tens{C}_0 \vec{\alpha}_i= \kappa_i$ before normalisation, then we immediately find $\vec{\alpha}_i \rightarrow \vec{\alpha}_i / \sqrt{\kappa_i}$ to satisfy  $\vec{\alpha}^{\dagger}_i \tens{C}_0 \vec{\alpha}_i=1$. Accordingly, the constant $d_i$ transforms as $d_i \rightarrow d_i/\sqrt{\kappa_i}$ as well, such that $\vec{\alpha}_i^{\dagger} \vec{\delta} =d_i$ is still obeyed after normalisation.\\
The consistency of the Gram-Schmidt condition $\vec{\alpha}^{\dagger}_i \tens{C}_0 \vec{\alpha}_j= \delta_{ij}$ and our normalisation $\vec{\alpha}_i^{\dagger} \tens{C}_{i-1}\vec{\alpha}_i =1$ can be shown as follows: Multiplying \eqref{app-eq-cov} by $\vec{\alpha}_{n}$ on both sides yields
\begin{align}
\vec{\alpha}_{n}^{\dagger} \tens{C}_{n-1} \vec{\alpha}_{n} &=  \vec{\alpha}_{n}^{\dagger} \tens{C}_0  \vec{\alpha}_{n} \nonumber \\ 
& \qquad - \sum_{i=1}^{n-1} \vec{\alpha}_{n}^{\dagger} \tens{C}_0 \vec{\alpha}_i \vec{\alpha}_i^{\dagger} \tens{C}_0 \vec{\alpha}_{n},
\end{align}
where we have shifted the index $n$ by 1 compared to \eqref{app-eq-cov} for ease of notation.
Inserting $\vec{\alpha}^{\dagger}_i \tens{C}_0 \vec{\alpha}_j= \delta_{ij}$ into the right hand side leads to
\begin{equation}
\vec{\alpha}_{n}^{\dagger} \tens{C}_{n-1} \vec{\alpha}_{n}= 1 - \sum_{i=1}^{n-1} \delta_{ni} \delta_{in} = 1,
\end{equation}
which proves the equality.

\section{Equivalence to HR91}
\label{app-B}
In this section we translate our notation into the one used by HR91 to show that our final expression \eqref{eq-app-final} is equivalent to their Eq.~(4), which states
\begin{equation}
\vec{\delta}_n= \vec{\delta}_0 + \sum_{i,j=1}^n \vec{\xi}_i \xi_{ij}^{-1} (d_i - d_{i0}),
\label{eq-app-HR}
\end{equation}
where we have already translated their notation for the constrained and unconstrained field, and the values of the constraints ($\vec{f} \rightarrow \vec{\delta}$, $c_i \rightarrow d_i$). The two additional functions are 
\begin{equation}
\vec{\xi}_i= \langle \vec{\delta}_0 \mathcal{C}_i^{\dagger} \rangle \\
\end{equation}
and
\begin{equation}
\xi_{ij}= \langle \mathcal{C}_i \mathcal{C}_j^{\dagger} \rangle,
\end{equation}
where $\mathcal{C}_i= \vec{\alpha}_i^{\dagger} \vec{\delta}_0$ (not to be confused with our covariance matrix $\tens{C}_i$), and we have added daggers to the HR91 notation to allow for complex-valued quantities. Inserting these expressions into \eqref{eq-app-HR} leads to
\begin{align}
\vec{\delta}_n&= \vec{\delta}_0 + \sum_{i,j=1}^n \tens{C}_0 \vec{\alpha}_i \left[ \vec{\alpha}_i^{\dagger} \tens{C}_0 \vec{\alpha}_j \right]^{-1}  (d_i - d_{i0}) \\
 &= \vec{\delta}_0 + \sum_{i=1}^n \tens{C}_0 \vec{\alpha}_i (d_i - d_{i0}),
\label{eq-app-HR2}
\end{align}
where in the first step we used that $\tens{C}_0 = \langle \vec{\delta}_0 \vec{\delta}_0^{\dagger} \rangle$ because $\vec{\mu}_0=0$ in HR91, and our orthonormalisation condition in the second step. This shows that the Gram-Schmidt approach is equivalent to performing the matrix inversion $\xi_{ij}^{-1}$.
\end{document}